\documentclass[superscriptaddress,preprintnumbers,showpacs,prb,aps,twocolumn]{revtex4-2}

\usepackage{amsfonts}
\usepackage{amssymb}
\usepackage{amsmath}
\usepackage{graphicx}
\usepackage{natbib}
\usepackage{physics}
\usepackage{siunitx}
\usepackage{float}
\usepackage{multirow}
\usepackage{ulem}
\usepackage{fancyhdr}
\usepackage{makecell}  

\usepackage[english]{babel}

\setcitestyle{numbers,square}

\usepackage[usenames,dvipsnames]{color}
\usepackage{soul}

\usepackage[hidelinks]{hyperref} 

\usepackage{tablefootnote}
\makeatletter
\newcommand\footnoteref[1]{\protected@xdef\@thefnmark{\ref{#1}}\@footnotemark}
\makeatother
\newcommand{\EF}{$E_\mathrm F$}
\newcommand{\rr}{$\sqrt(13)\times\sqrt(13)R13.9^\circ$}

\newcommand{\micron}{$\mathrm{\mu}$m}
\newcommand{\Kxx}{$\kappa_{\rm {xx}}$}
\newcommand{\Kxy}{$\kappa_{\rm {xy}}$}

\newcommand{\hb}{4$Hb$-Ta(S$_{1-x}$Se$_{x})_2$}
\newcommand{\Fhb}{4$Hb$-TaS$_2$}
\newcommand{\Thb}{2$H$-TaS$_2$}

\newcommand{\onet}{1$T$-TaS$_{2-x}$Se$_x$}
\newcommand{\oneh}{1$H$-TaS$_{2-x}$Se$_x$}

\newcommand{\moire}{Moiré}

\newcommand{\el}{$e^-$}
\newcommand{\ua}{$\mu$ARPES}

\newcommand{\Ts}{$T$-surface}
\newcommand{\Hs}{$H$-surface}
\newcommand{\Tl}{$T$-layer}
\newcommand{\Hl}{$H$-layer}
\newcommand{\G}{$\Gamma$}
\newcommand{\vs}{\textit{vs}.}
\newcommand{\cc}{$\overline{\mathrm{CC}}$}
\newcommand{\ccmath}{\overline{\mathrm{CC}}}
\newcommand{\dzz}{$d_{z^2}$}
\newcommand{\dxx}{$d_{x^2-y^2}$}
\newcommand{\dxz}{$d_{xz}$}
\newcommand{\dxy}{$d_{xy}$}
\newcommand{\dyz}{$d_{yz}$}

\newcommand{\tbt}{$3\times3$}
\newcommand{\rthirteen}{$\sqrt{13}\times\sqrt{13}\mathrm{R}13.9^{\circ}$}
\newcommand{\rthirt}{$\sqrt{13}$}

\newcommand{\dW}{$\Delta W$}
\newcommand{\zpi}{0\%$I$}
\newcommand{\zpm}{0\%$M$}
\newcommand{\opm}{1\%$M$}
\newcommand{\qpm}{0.25\%$M$}
\newcommand{\otts}{$1T-$TaS$_2$}

\newcommand{\ky}{$k_y$}

\begin{document}

\title{Bulk Superconductivity driven by Disorder-Induced Delocalization in \hb}

\author{Lu~Chen}
\thanks{Equal contribution}
\affiliation{Materials Science Division, Lawrence Berkeley National Laboratory, Berkeley, CA 94720, USA}
\affiliation{Department of Physics, University of California, Berkeley, California 94720, USA}
\affiliation{Department of Physics, University of Illinois Urbana-Champaign 61801, USA}

\author{Sae-Hee Ryu}
\thanks{Equal contribution}
\affiliation{Advanced Light Source, Lawrence Berkeley National Laboratory, Berkeley, CA 94720, USA}

\author{Avior Almoalem}
\affiliation{Department of Physics, University of Illinois Urbana-Champaign 61801, USA}

\author{Yuanqi~Lyu}
\affiliation{Department of Physics, University of California, Berkeley, California 94720, USA}

\author{Koh~Yamakawa}
\affiliation{Department of Physics, University of California, Berkeley, California 94720, USA}

\author{Luke~Pritchard~Cairns}
\affiliation{Department of Physics, University of California, Berkeley, California 94720, USA}

\author{Ryan~Day}
\affiliation{Department of Physics, University of California, Berkeley, California 94720, USA}

\author{Ehud Altman}
\affiliation{Department of Physics, University of California, Berkeley, California 94720, USA}

\author{Daniel Podolsky}
\affiliation{Physics Department, Technion, Haifa 32000, Israel}

\author{Dung-Hai Lee}
\affiliation{Advanced Light Source, Lawrence Berkeley National Laboratory, Berkeley, CA 94720, USA}

\author{Vidya Madhavan}
\affiliation{Department of Physics, University of Illinois Urbana-Champaign 61801, USA}

\author{Eli Rotenberg}
\affiliation{Advanced Light Source, Lawrence Berkeley National Laboratory, Berkeley, CA 94720, USA}

\author{James~G.~Analytis}
\email{luchen721@berkeley.edu, analytis@berkeley.edu}
\affiliation{Materials Science Division, Lawrence Berkeley National Laboratory, Berkeley, CA 94720, USA}
\affiliation{Department of Physics, University of California, Berkeley, California 94720, USA}
\affiliation{CIFAR Quantum Materials, CIFAR, Ontario, Toronto M5G 1M1, Canada}
\affiliation{Kavli Energy NanoScience Institute, Berkeley, California 94720, USA}

\date{\today}

\pagestyle{plain}

\begin{abstract}

 The unconventional superconductor \Fhb~is a natural heterostructure that can be broadly understood as interleaving Mott-like and metallic layers. We study the properties of this material as a function of quenched disorder in the form of Se/S substitution and find that while disordered samples show bulk superconductivity, clean samples do not. We show that a disorder-driven delocalization of carriers in the Mott-like ($1T$-) layer forms a new Fermi surface that is absent in the cleanest samples. This suggests that one of the primary drivers for superconductivity is the fragility of the flat band, whose delocalization brings to life a sea of strongly correlated electrons.

\end{abstract}

\pacs{Valid PACS appear here}

\maketitle

Delocalization transitions underlie many important problems in condensed matter physics, particularly unconventional superconductivity. The observation is that a system undergoes a small to large Fermi surface change as a function of some non-thermal parameter and this often coincides with the highest critical temperatures. The implication is that understanding the mechanism behind the delocalization will point to the mechanism of superconductivity.\cite{paschen_hall-effect_2004,badoux_change_2016, maksimovic_evidence_2022} We explore this question in the unconventional superconductor~\Fhb, and show that there is indeed a Fermi surface transition in a flat band that is correlated with the appearance of bulk superconductivity. However, there is a surprising origin to this delocalization: disorder. 


TaS$_{2}$ is an exemplary system where different lattice structures can host very different quantum ground states \cite{Wilson1974,Balseiro1979, Sipos2008}; it forms in multiple structural polytypes $1T$, $2H$, or \Fhb,  manifests various charge density wave (CDW) orders \cite{Wilson1974}, superconductivity \cite{Wilson1974}, Mott physics \cite{Wang2020}, and potentially a quantum spin liquid phase \cite{Law2017,Yu2017}. In such a system, structural disorder should be expected to have a significant effect on the quantum order of the system.

\begin{figure*}[t]
\centering
\includegraphics[width = 1 \linewidth]{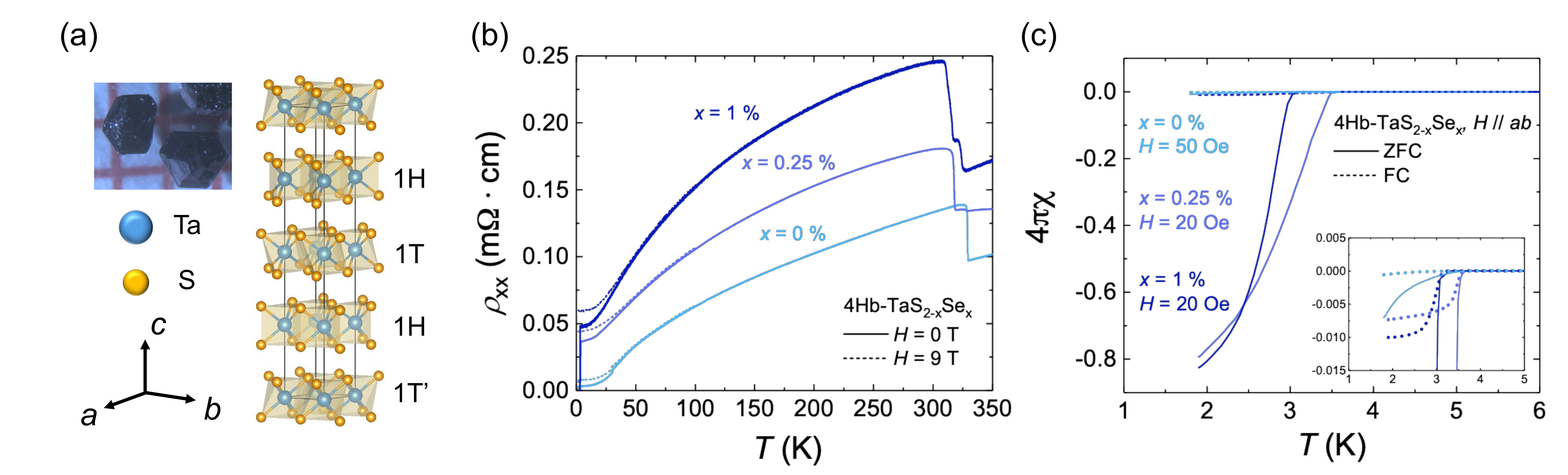}
\caption{{\bf Sensitivity of superconductivity to Se substitution.}
(a) Crystal structure of \hb. \onet~and \oneh~layers stack along the crystal $c$ axis with van der Waals coupling between the layers. Single crystals of \hb~are also shown. (b) Temperature dependence of the in-plane electrical resistivity $\rho_{xx} (T)$ for \hb~single crystals with three different Se concentrations at magnetic fields of $H$ = 0 T (solid line) and $H$ = 9 T (short dashed line). Only the $x=0 \%$ sample shows a kink in resistivity around $T\sim30$ K due to the appearance of a 3 $\times$ 3 commensurate charge density wave (CCDW) in the 1H layer. This low-temperature CCDW transition is suppressed in the $x=0.25 \%$ and $x=1 \%$ samples.
(c) Temperature dependence of the magnetic susceptibility in three \hb~single crystals with different Se concentration: $x=0 \%$, $x=0.25 \%$, and $x=1 \%$. Inset shows a zoom-in view of $4\pi\chi(T)$ curves. {The superconducting volume fraction (SCVF) is only 0.6\% for the $x=0 \%$ sample, while it is greatly enhanced to about 80\% for the $x=0.25 \%$ and $x=1 \%$ samples.}
}
\label{Fig:transport_mag}
\end{figure*}

Compared to the $1T$- and $2H$- polytypes of TaS$_{2}$, \Fhb~has attracted tremendous attention recently because its structure represents a natural interleaving of the $1T$ and $1H$ layers (as shown in Fig. \ref{Fig:transport_mag} (a)), combining key structural and electronic features of both a Mott insulator and a superconductor \cite{Salvo1973}. {The former provides a localized `flat band' subsystem, while the latter provides a sea of itinerant charge carriers, meeting in three-dimensionally stacked interfaces.} Superconductivity with a transition temperature in the range of 2.7 - 3 K has been observed, and associated with a number of unconventional properties including time reversal symmetry breaking \cite{Ribak2020}, topological boundary modes \cite{Nayak2021}, spontaneous vortices \cite{Persky2022}, and multi-component order parameters \cite{Silber2024,Almoalem2024} have been reported recently.

 However, the connection of the superconducting order to the parent Mott and metallic normal state remains opaque. In this study we show that the presence of superconductivity in \Fhb~ is strongly connected to disorder-driven delocalization of carriers in the \Tl, while the suppression of the CDW in the \Hl\ is, by comparison, a relatively small effect. Superficially, this transition is reminiscent of a small to large Fermi surface transition seen in heavy fermion superconductors \cite{paschen_hall-effect_2004, maksimovic_evidence_2022}. We suggest that it is the fragility of the flat band that is the primary culprit, and this allows small amounts of disorder to liberate carriers and creating a new Fermi surface that facilitates superconductivity.
 
 \section{Results}

    Selenium is isovalent with sulfur, and so the primary effect of Se substitution should be to create local disorder in the material without substantially changing the carrier density. Our evidence, shown in Supplementary Materials Sec.~\ref{sec:corelevel}, indicates Se defects behave well in \hb, substituting randomly for S in both $T$ and $H$ layers and not taking unusual interstitial positions, clustering, or unexpected chemical bonding. In Figure \ref{Fig:transport_mag} (b), we illustrate the resistivity curves for three samples of \hb, at $x$ = 0\%, 0.25\% and 1\%. The trend is very typical of adding disorder in pristine samples in that the resistivity increases monotonically with Se concentration.

    The pristine samples have very long mean free paths, which can be observed in the residual resistivity ratio which evolves by more than an order of magnitude from $\rho_{300}/\rho_{T_{c+}}=66$ at $x$ = 0\% to $\rho_{300}/\rho_{T_{c+}}=5$ at $x=1\%$, where $T_{c+}$ is defined as the normal state temperature just above $T_c$. The latter residual resistivity ratio is much more typical of samples presented in the literature \cite{Gao2020,Meng2024}, and the high mobility samples presented here ($x$ = 0\%) will play an important role in understanding the underlying physics of these materials. Importantly, all samples show evidence for the incommensurate CDW (ICCDW) appearing above room temperature, which is known to be the Star-of-David (SOD) ordering with approximately \rr\ (henceforth \rthirt) symmetry in the 1T layers \cite{meyer_properties_1975}.

The magnetization shows its own peculiarities. As shown in Fig. \ref{Fig:transport_mag} (c), the onset of superconductivity occurs at temperatures as high as $T_c=3.5$ K for the $x$ = 0.25 \% sample, whereas the $x$ = 1 \% sample shows an onset at $\sim 3$ K. However, the most important observation is the striking contrast of the volume fraction with the pristine samples. These data show a vanishingly small volume fraction of superconductivity, less than $0.6\%$, whereas just a small amount of Se brings up the volume fraction to $80\%$ (Given the demagnetization factor of the sample, the volume-metric fraction in the Se-substituted sample is unitary). Small amounts of disorder increase the volume fraction by more than two orders of magnitude, a fact that has been noticed before.\cite{Meng2024}

\begin{figure}[t]
\centering
\includegraphics[width = 1.0 \linewidth]{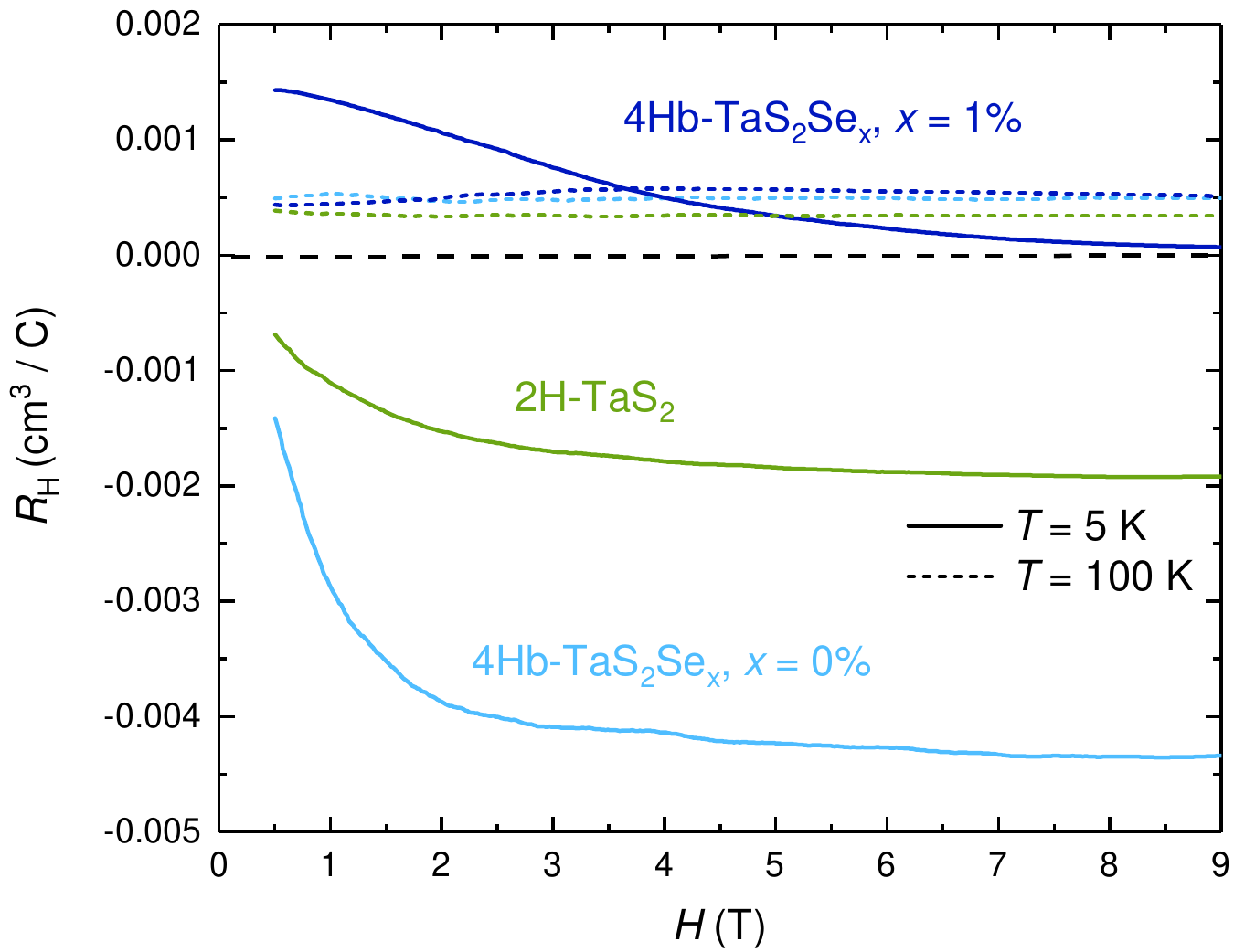}
\caption{{\bf Evidence for delocalization transition in \hb\, as a function of Se substitution.} Hall coefficient $R_{\mathrm{H}}$ in the high field limit at low temperature (5 K, solid lines) and high temperature (100 K, dashed lines) measured in 2H-TaS$_{2}$ (green curves) and \hb~samples at $x$ = 0\% (light blue curves) and 1\% (dark blue curves). At all temperatures the \hb~$x$ = 0\% sample has roughly half the carriers of 2H-TaS$_2$, consistent with all the carriers in the 1T layers being localized. At low temperatures the \hb\, at $x$ = 1\% sees a dramatic change where the Hall coefficient not only changes sign, but asymptotes to small value of R$_H$, suggesting a huge increase in carrier density.}
\label{Fig:4HbHall}
\end{figure}
The Hall coefficient $R_H(H)$ as a function of magnetic field measured in $2H$-TaS$_{2}$ and \hb~samples at $x$ = 0\% and 1\% are shown in Figure \ref{Fig:4HbHall}. The field dependence of in-plane electrical Hall resistivity $\rho_{xy}(H)$ for \hb~$x=0 \%$, $x=0.25 \%$, and $x=1 \%$ samples measured at selected temperatures are presented in the Supplementary Materials Sec.~\ref{sec:hall}. The Fermi surface of the \hb~system is quite complicated, making the interpretation of the Hall number challenging. However, the high field limit is well defined in any multi-band system, giving the difference between the number of electrons and the number of holes, i.e. $R_H(\infty)=-1/e(n_e-n_h)$. Experimentally, this regime is typically identified by a saturation in $R_H(H)$, flat-lining at $R_H(\infty)$. At high temperatures ($T\sim$ 100 K), the samples are very well behaved, with $x$ = 0\% and 1\% showing an essentially unchanged Hall number. Interestingly, $R_H(\infty)$ for \hb~is just under double that of $R_H(\infty)$ for the $2H-$TaS$_2$, consistent with the expected localization of the 1T layers of \hb~which account for half of the structure. 

At lower temperatures ($T\sim$ 5 K), the data have more structure. The pristine \Fhb~sample continues the expected trend, exhibiting a value of $R_H(\infty)$ that is approximately twice that of the $2H-$ structure and of the same sign. However, as soon as a small amount of Se is added, something dramatic happens: not only is there a sign change in the Hall coefficient, but the approach to the high-field limit changes qualitatively and quantitatively. (The $2H$-TaS$_2$ has been reported to show similar behavior as a function of thickness, which may also be related to disorder/strain.\cite{yang_enhanced_2018}) For $x$ = 1\% sample, the Hall number is large at low fields and decreases at high fields, approaching a value corresponding to a significantly higher carrier density. However, we do not reach the high-field limit, emphasizing that the Fermi surfaces are indeed very different from the pristine case. For $x$ = 0\% we observe $n_e-n_h\approx 2\times10^{21}$cm$^{-3}$. 

\begin{figure}[b]
\centering
\includegraphics[width = 1 \linewidth]{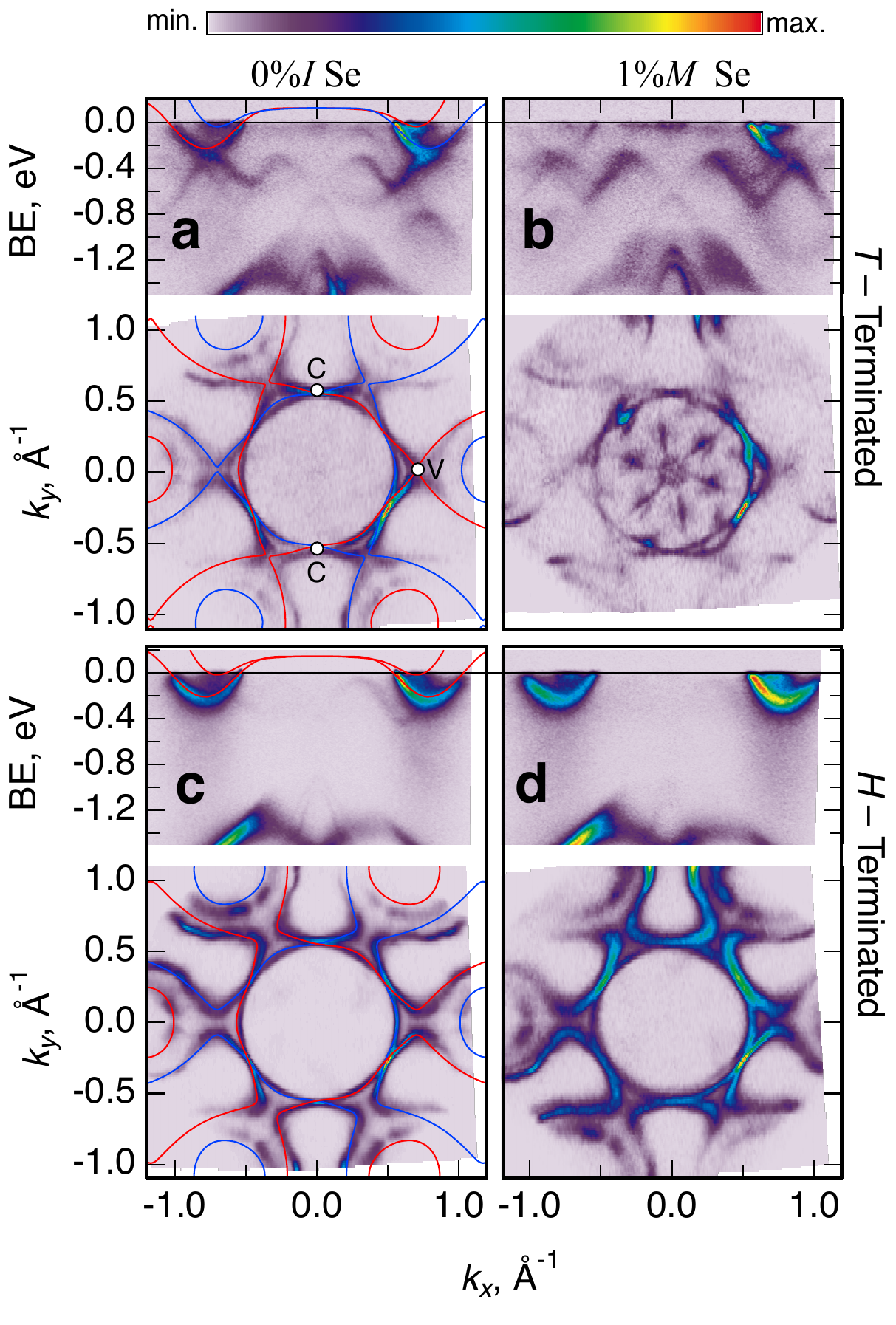}
\caption{{\bf ARPES evidence for a delocalization transition in \hb.} Slices of the full ARPES Binding Energy (BE) \vs\ $\mathbf{k}$\ data set for (a,b) 1$T$-terminated and (c,d) 1$H$-terminated regions of the sample surface. The upper part of each panel is a bandstructure slice (BE \vs\ $k_x$ at $k_y=0$, while the lower part is the intensity distribution  at \EF~along $k_x$ and $k_y$.  The ARPES data shows the same dramatic effect as the electrical Hall effect: there is a filling of the CDW gap in the 1$H$ layers (red circles) while in the 1$T$ layer a metallic Fermi surface appears near \G\ with 1$\%$ Se substitution.   The data are collected at photon energy 72 eV, and all panels use the same color scale linearly mapped onto the individual intensity ranges.
}
\label{Fig:ARPES}
\end{figure}

Although small quantities of defects do not, in general, significantly affect the Fermi surface of metals, the dramatic changes we observe are difficult to explain otherwise. To gain further insight, we turn to spatially-resolved ARPES \cite{Ribak2020,Almoalem2024ARPES,Date2025, Gofman2025,Li2025,Sun2025,Watson2025,Yang2025,YangFZ2025, Li2025}, a probe which is directly sensitive  to interlayer charge transfer, lattice symmetries, orbital order, and band renormalization due to interactions. Fig.\ \ref{Fig:ARPES} (a) shows the \ua\ electronic structure of $T$-terminated surfaces for \Fhb~with 0\% Se. Although not evident at \EF, replica bands observed at higher binding energy prove that the \Tl\ for  all samples presented is in the \rthirt\ reconstructed phase associated with the CDW formation and SOD cluster arrangement. Overlaid on the data is a tight-binding (TB) model for an isolated \Hl~(see Supplementary Materials Sec. \ref{sec:Hlayer}). Its close match to the measured bands establishes that the Fermi surface observed on the $T$-terminated surface of the 0\% sample exhibits only the features associated with the underlying \Hl, demonstrating that the surface \Tl\ is insulating.  We denote this class of samples as ``\zpi''.

Fig.\ \ref{Fig:ARPES}(b) shows the \Ts\ Fermi surface for 1\% Se-doped samples. {In addition to the H-layer-derived band features exhibited by the \zpi\ samples in (a), a new set of sharp metallic ``windmill'' states appears around \G.} Although recent studies have attributed these chiral windmill features directly to the \Tl\ flat band \cite{Date2025,Gofman2025}, their dispersion indicates they are actually replica bands of the underlying \Hl\ folded by the \rthirt\ lattice superpotential \cite{Li2025}. Crucially, these itinerant \Hl\ replica states hybridize with the \Tl\ flat band, which has been interpreted previously as a metallic Kondo-like resonance at \EF\ \cite{Li2025, Shen2022}. Similar states are apparent in samples with 0.25\% Se concentration (not shown here). Because these $T$-surfaces are metallic, the Se-doped samples are designated as \opm\ and \qpm, respectively. Ultimately, the emergence of this new, delocalized Fermi surface on the \Tl\  accounts for the dramatic changes observed in the size and amplitude of the high-field Hall effect.

Fig.\ \ref{Fig:ARPES}(c,d) show the Fermi surfaces of the \Hs\ layers for the different sample classes. All $H$ layers show signatures of \tbt\ CDW reconstruction, evidenced by strongly reduced intensity along the hole pockets around the K points and the ``dog-bone'' pockets near the M points of the Brillouin Zone. Quantitative analysis of the \Hl\ Fermi-surface areas reveals a finite charge transfer from the $T$ to $H$ layers, but our extracted values remain substantially smaller than the 1\,\el\ per SOD that would be required to fully empty the \Tl\ flat band (full details in Supplementary Materials Sec.\ \ref{sec:Hlayer}). Specifically, for a bulk \Tl\ sandwiched between two 1$H$ layers, the retained charge is $\sim 0.46$\,\el/SOD in \zpi\ samples, rising to $\sim 0.70$\,\el/SOD with 1\% Se --- in both cases the flat band retains the majority of its charge. These results directly contradict a complete charge-transfer scenario and confirm that the \Tl\ remains partially occupied across all sample classes. {It then appears that Se induces a transition in this state which delocalizes the flat band, enabling} the hybridization of electrons between layers, manifested by the appearance of a new Fermi surface on the \Tl.
 
\begin{figure}[t]
\centering
\includegraphics[width = 1 \linewidth]{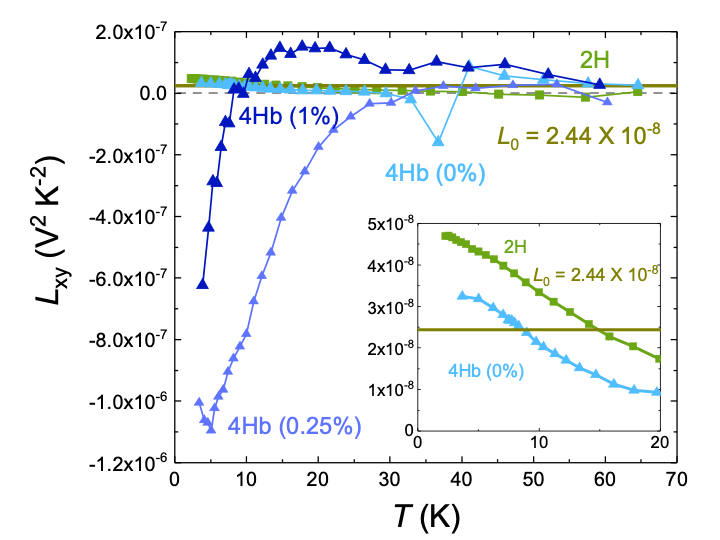}
\caption{{\bf Lorenz number from the thermal Hall conductivity in \hb.}
Calculated Lorenz number $L_{xy} = \kappa_{xy}/(T\sigma_{xy})$ in 2H-TaS$_{2}$, \hb~$x$ = 0\%, $x$ = 0.25\%, and 1\% samples in a magnetic field of $H = 9$~T. The Wiedemann-Franz law is mostly fulfilled in 2H-TaS$_{2}$ and pristine \hb (as shown in the inset panel), while it is strongly violated in the Se substituted \hb~samples.}
\label{Fig:4HbthermLxy}
\end{figure}

To understand the causal connection between the Fermi surface transition and disorder, we study the bulk thermal conductivity (which is sensitive to lattice disorder), and local scanning tunneling microscopy. Most of the thermal conductivity data is discussed in Supplementary Materials \ref{sec:kappaxx}, and is consistent with a strong increase in lattice scattering with increasing Se. The most striking effect is found in studying the Wiedemann-Franz law in the thermal Hall effect, which states that in a conventional metal the {Hall Lorenz ratio $L_{xy} =\kappa_{xy}/(\sigma_{xy}T)$ approaches the Sommerfeld value $L_0$.} Notably, the \zpi\ samples look very consistent with the Wiedemann-Franz law at low temperatures, as do 2H-TaS$_2$ samples, consistent that most of the conduction arises from \Hl~{(as shown in Fig. \ref{Fig:4HbthermLxy})}. However, in \opm and \qpm~samples the Wiedemann-Franz law is strongly violated below the $3\times 3$ CDW transition. This suggests additional multi-band effects come in to play as the \Tl\ and \Hl\ becomes coupled. The violation of Wiedemann-Franz is therefore not explicit evidence of Mott delocalization but of changes to the electronic structure.

\begin{figure}[t]
\centering
\includegraphics[width = 0.5\textwidth]{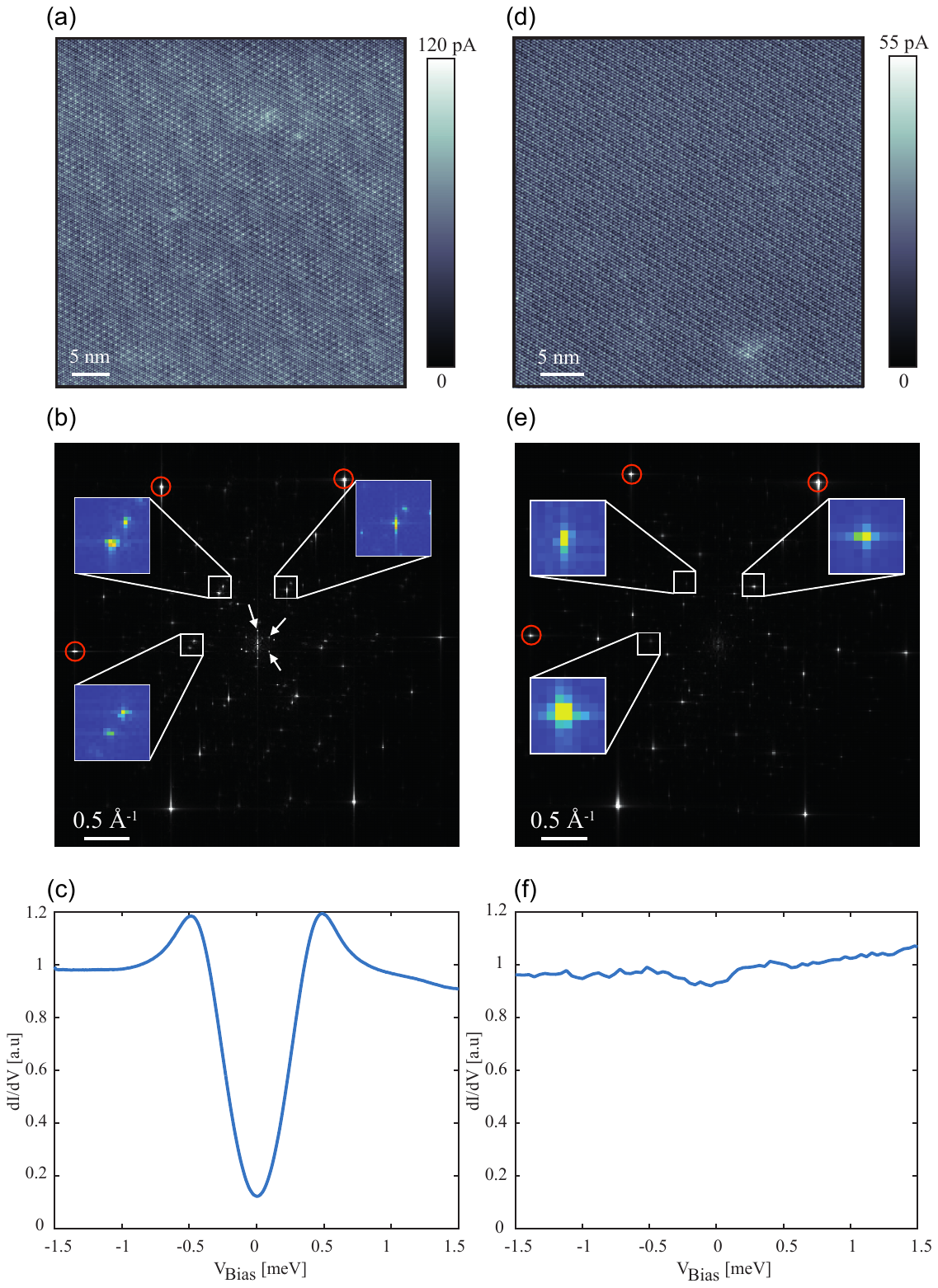}
\caption{
{\bf Correlation between the SC-gap and CDW domains in pristine \Fhb.} (a) Atomically resolved topography of the $1H$ layer ($V = -5$mV, $I = 70$mV). (b) Fast Fourier transform (FFT) of the topography in (a), with white arrows indicating central moir\'e peaks. (c) Averaged tunneling spectrum acquired on the surface shown in (a), displaying a well-defined superconducting gap ($V_\mathrm{set}$ = -1.5 mV, $I$ = 70 pA, $V_\mathrm{modulation}$ = 30 $\mu$eV). (d, e) Topography and corresponding FFT at a different location showing unsplit CDW peaks. (f) Averaged spectrum in the region shown in (d), indicating a highly suppressed gap ($V_\mathrm{set}$ = -2 mV, $I$ = 100 pA, $V_\mathrm{modulation}$ = 40 $\mu$eV). Spectra in (c) and (f) were taken with the same tip on the same sample. All measurements were obtained at $T \sim$ 300 mK.}
\label{Fig:STM}
\end{figure}

Our STM studies of twinned and non-twinned regions of \Hl\ CDW domains of \zpi\ samples confirm that the appearance of superconductivity is correlated with disorder. Figure \ref{Fig:STM} (a,d) shows the atomically resolved topography obtained on the \Hl\ and fast Fourier Transform (FFT) in Figure \ref{Fig:STM} (b,e) for twinned and non-twinned regions respectively. In the former, a CDW state is formed with multiple domains, manifested in real space as a local $\sim 3\times 3$ order having a prominent stripe modulation running from lower left to upper right and period $\sim 7$ nm.  This modulated CDW appears in $k$-space by split peaks near the $3\times 3$ Bragg peaks in \ref{Fig:STM} (b) \cite{Yan2017,Wang2020Science}, which is absent in non-twinned regions, Fig. \ref{Fig:STM} (e). Additional, weaker Bragg peaks, most notably a central hexagon ring (white arrows), in the FFT correspond to additional weaker \moire\ modulations of the CDW at scales $> 4$ nm that are observable in the STM map due to the \Tl immediately below. These appear to be suppressed in the non-twinned regions. Fig. \ref{Fig:STM}(c,f) shows a striking difference in the superconducting signatures between the two regions: while twinned regions have a well-defined superconducting gap, this is strongly suppressed in areas with a single CDW domain. We do not have equivalent data on \Tl s, but this does not affect the broad observation: the only regions where superconductivity can be found in \zpi\ samples, are regions where there is disorder{, consistent with the negligible ($<0.6\%$) volume fraction of superconductivity.}

\section{Discussion}

Small Se substitution is important for making \Fhb\ a bulk superconductor. The lightest compositions $x\sim0.25\%$ show evidence for an onset at $\sim$3.5 K, the highest reported $T_c$. At this concentration the Se-Se distance averages about 6 nm, strikingly similar to the length scale of disorder-induced superconductivity observed in STM in Se-free samples, Fig. \ref{Fig:STM}. This coincidence of length scales may be a clue as to how Se disorder (existing throughout the bulk) leads to bulk superconductivity.


Theoretical models predict the flat band on the \Tl\ to be half-filled and derived from Ta \dzz\ orbitals localized at the center of the SOD cluster, driving the system into a strongly correlated flat band insulating state \cite{Fazekas1980,Wang2020,Petocchi2022}. This intrinsic correlation-driven state has been confirmed by recent experimental observations of isolated monolayer \otts, which demonstrate a robust 2D Mott gap \cite{chen_strong_2020,lin_scanning_2020}. Recently, it was proposed that massive interlayer charge transfer completely depletes this flat band in the heterostructure, effectively destroying the Mott physics \cite{Almoalem2024ARPES, Date2025}. However, our data explicitly contradicts this: in our pristine \zpi\ samples, the 1\Tl's flat band is not completely depleted, but remains at least partially occupied. This is evidenced by diffuse states that tail up to \EF\ with a maximum intensity around \EF-25 meV. As detailed in Supplementary Materials Sec.\ \ref{sec:polarization}, our polarization-dependent ARPES measurements explicitly confirm the \dzz\ orbital character of these localized states, directly linking them to the persistent \Tl\ flat band.

{We use the term \textit{hybridization} in a generic sense: while prior work has discussed the flat-band/itinerant-band coupling in terms of heavy-fermion or Kondo states\cite{Li2025}, implying spectral features that our data do not resolve, we make no specific claim about the microscopic screening mechanism.  We note that evidence for Kondo or heavy-fermion behaviour has been discussed theoretically\cite{Lin2024,Konig2024}, and experimentally by STM measurements on \Fhb\cite{Shen2022,Nayak2023} and related heterostructures\cite{Vano2021,Fei2022,gonzalez_thesis}.
}

We propose that weak Se substitution destabilizes the partially localized $1T$- flat band state by enhancing bandwidth and charge fluctuations, thereby driving a correlation-induced delocalization transition. There are two likely cooperating correlation mechanisms for Se's outsized effect on the \Tl. First, Se introduces significant $p$-$d$ hybridization between its Se~$4p$ and Ta~$5d$ orbitals that is absent when S occupies the chalcogen site \cite{Bovet2004,Ang2013}; this increases the effective \Tl\ bandwidth $W$ and reduces the ratio $U/W$ governing the Mott-like physics \cite{Ang2013}, directly enabling the delocalization transition evidenced by ARPES, the Hall coefficient, and the violation of the Wiedemann-Franz law. Second, once the \Tl\ becomes metallic, there is substantial energy to be gained through hybridization of the itinerant \Tl\ states with the \Hl\ replica bands folded by the \rthirt\ superpotential, opening gaps across large regions of k-space. This condensation energy competes directly with the electrostatic driving force for $T\rightarrow H$ charge transfer, further stabilizing the occupied flat band and making the metallic hybridized ground state self-reinforcing once reached. It is possible that this second mechanism is present with all kinds of disorder capable of destabilizing the flat band in the \Tl, even in nominally pure sulfur compounds.

The remarkable sensitivity to Se substitution does not preclude competition between the commensurate CDW and superconducting order parameters as previously suggested \cite{Ribak2020, Meng2024}, but neither does it require it. Rather, it is the fragility of the flat band that liberates carriers that can then become superconducting, while the CDW in the \Hl\ experiences a much weaker effect with only minor changes to the Fermi surface. This means that native disorder can similarly destroy the flat band, and so even samples sans Se may exhibit signatures of superconductivity, albeit with a volume fraction that depends on the extent of disorder (see suppl.~\S~IV.D).\cite{Meng2024} It is interesting that early experiments on \Fhb\ considered samples of substantially higher quality than those generally studied today, and superconductivity does not appear to be present.\cite{fleming_oscillatory_1977}

\section{ACKNOWLEDGEMENTS}
This work was primarily funded by the U.S. DOE, Office of Science, Office of Basic Energy Sciences, Materials Sciences and Engineering Division under Contract No. DE-AC02-05CH11231 (Quantum Materials Program KC2202). S-H. R. and E. R. work was part of the QSA, supported by the U.S. Department of Energy, Office of Science, National Quantum Information Science Research Centers. This research used resources of the Advanced Light Source, which is a DOE Office of Science User Facility under contract no. DE-AC02-05CH11231.
We thank Chris Jozwiak and Aaron Bostwick for stimulating discussions and assistance with beamtime at the ALS.  

\section{METHODS}

\subsection{Samples}
\hb~single crystals are synthesized by chemical vapor transport (CVT) method with iodine as a transport agent \cite{Salvo1973}.
Stoichiometric amounts of Ta (Alfa Aesar 99.98\%), S (ThermoFisher Scientific 99.999\%), and Se (ThermoFisher Scientific 99.999\%) powders are mixed and sealed inside a quartz ampoule under vacuum.
The precursor is synthesized by heating up the mixture at 900 $^{\circ}$C for 84 hours and then slowly cool down to room temperature.
Then the obtained precursor and iodine are mixed, ground, and sealed inside a quartz ampoule under vacuum.
The mixture is heated for 27 days in a two-zone furnace,
where the temperature of the source zone and growth zone are fixed at 780 $^{\circ}$C and 680 $^{\circ}$C,
then cooled down with the furnace turned off.
This process yields hexagonal-shaped bulk crystals with naturally formed sharp edges as illustrated in Fig.\ref{Fig:transport_mag} (a) with a typical length of 1 to 2 mm.

1T-TaS$_{2}$ single crystals are synthesized using CVT method with elemental Ta, and S in a stoichiometric ratio of 1 : 2. The powder is mixed and ground together, then loaded into an alumina crucible with iodine and sealed under vacuum in a quartz tube. The mixture is placed in a two-zone furnace with both zones heated up to 950 $^{\circ}$C for 6 hours to encourage nucleation. Then, the source zone is raised to
1050 $^{\circ}$C while the growth zone is kept at 950 $^{\circ}$C for 7 days. After that, the quartz  tube is quenched in ice water, and 1T-TaS$_{2}$ single crystals are harvested.

2H-TaS$_{2}$ single crystals are synthesized using CVT method with elemental Ta and S in a stoichiometric ratio of 1 : 2. The precursor is synthesized by mixing and grinding Ta and S powder, which are then loaded into an alumina crucible and sealed in a quartz tube under a partial pressure of argon. The sealed tube is heated to 400 $^{\circ}$C and held for 6 hours, then ramped to 900 $^{\circ}$C and held  for 10 days. The resulting precursor is subsequently sealed under vacuum with iodine
in a quartz tube and placed in a two-zone furnace, with both zones initially held at 850 $^{\circ}$C for 6 hours to promote nucleation. Then, the source zone is increased to
950 $^{\circ}$C, while the growth zone is kept at 850 $^{\circ}$C for 10 days. Finally, the furnace is shut off and allowed to cool naturally to room temperature.

For electrical and thermal transport measurements, 
the bulk crystals are further thinned down to $\sim100$ $\mu m$ by exfoliation using Scotch tape.
Contacts are made using silver paste and silver wires. 
The dimensions (length between contacts $L$ $\times $ width $w$ $\times$ thickness $t$, in $\mu$m) of all the measured samples are listed in TABLE I. 

\setlength{\tabcolsep}{4pt}
\begin{table}[!]
  \centering
  \label{Table:thermal_conductivity}
\begin{tabular*}{\linewidth}[t]{ccccc}
    \hline
    \hline
    \noalign{\vskip 0.1cm} 
   Material & Se \% & $L$ ($\mu$m) & $w$  ($\mu$m)   & $t$  ($\mu$m)   \\
    \hline
    \noalign{\vskip 0.1cm}
4Hb-TaS$_{2}$  & $x$ = 0 \% & 281 & 573 & 63 \\

4Hb-TaS$_{1.995}$Se$_{0.005}$  & $x$ = 0.25 \% & 358 & 476 & 124  \\

4Hb-TaS$_{1.98}$Se$_{0.02}$  & $x$ = 1 \% & 513 & 1101 & 140 \\

1T-TaS$_{2}$  & $x$ = 0 \% & 1107 & 1109 & 94 \\

2H-TaS$_{2}$  & $x$ = 0 \% & 723 & 1538 & 15  \\
    
    \noalign{\vskip 0.1cm}
    \hline
    \hline

\end{tabular*}
\caption{Sample information including the Se concentration level and dimensions of the contacts (length between contacts $L$ $\times $ width $w$ $\times$ thickness $t$, in $\mu$m).
The Se concentration level $x$ is defined as the number of Se atoms added to per unit formula of 4Hb-TaS$_{2}$.
}
\end{table}

\subsection{Magnetization and Transport Measurements}

The dc magnetic susceptibility ($\chi$) is measured using a Quantum Design Superconducting Quantum Interference Device (SQUID) system.
The in-plane electrical resistivity ($\rho_{xx}$) and electrical Hall resistivity ($\rho_{xy}$) are measured by the standard four-probe method with the electrical current in the $ab$ plane and magnetic field along $c$ axis in a Quantum Design Physical Property Measurement System (PPMS).

The thermal conductivity \Kxx~and thermal Hall conductivity \Kxy~are measured in the same samples where $\rho_{xx}$ and $\rho_{xy}$ are measured by applying a heat current $J$ inside the $ab$ plane (along the naturally formed edges of the hexagonal shape) and a magnetic field $H$ along the crystal $c$ axis.
We define the $J$ direction to be $x$ and the $H$ direction to be $z$.
The heat current $J$ generates a longitudinal temperature difference $\Delta T_{\rm{x}} = T^{+} - T^{-}$ along $x$.
Then the thermal conductivity \Kxx~can be defined by
\begin{align}
  \kappa_{\rm{xx}} = \frac{J}{\Delta T_{\rm x}}\left(\frac{L}{wt}\right),
\end{align}
where $w$ is the sample width, $t$ its thickness and $L$ the distance between $T^{+}$ and $T^{-}$.
The transverse temperature difference $\Delta T_y$ is measured along the $y$ direction.
The thermal Hall conductivity \Kxy~is then given by 
\begin{align}
  \kappa_{\rm{xy}} = -\kappa_{\rm{yy}}\left(\frac{\Delta T_{y}}{\Delta T_{x}}\right)\left(\frac{L}{w}\right).
\end{align}
In this study, we use $\kappa_{\rm{xx}} = \kappa_{\rm{yy}}$ as an approximation.
More detailed discussions on crystal symmetry and an accurate measurement of $\kappa_{\rm{yy}}$ in a hexagonal shaped sample with a six-fold rotational symmetry can be found in Ref. \cite{Chen2024,Chen2024PRX}.

The thermal conductivity \Kxx~and thermal Hall conductivity \Kxy~are measured in a PPMS using a home-built thermal transport puck.
The thermal transport measurement is carried out by a steady-state method using a one-heater and three-thermometer configuration.
A resistive heater is connected to one end of the sample to provide a thermal gradient.
The other end of the sample is mounted on the copper heat sink using silver paint.
Both the longitudinal and transverse temperature difference $\Delta T_x$ and $\Delta T_y$ are measured using Cernox thermometers.

The measurement sequence is described as follows:
a magnetic field of $+H$ is applied at 70 K, 
then the sample is cooled down to the base temperature of 2 K with a $+H$ field.
The $+H$ data is taken by changing temperature in discrete steps under $+H$ field during warm up.
At each temperature step, 
after the temperature is stabilized,
the background is eliminated by subtracting the readings at the heater-off status from the heater-on status.
After the $+H$ run is finished,
a magnetic field of $-H$ is applied at the same temperature as the $+H$ run, i.e. at 70 K.
Then the $-H$ run is carried out with the same procedure.
The Cernox thermometers are calibrated in-situ using the heater-off readings at each temperature step.
The contamination from \Kxx~in \Kxy~due to contact misalignment is removed by doing field anti-symmetrization of the transverse temperature difference $\Delta T_y$. 
That is to say, $\Delta T_y$ are measured with both positive and negative magnetic fields in the same conditions, then \Kxy~is calculated using the field anti-symmetrized $\Delta T_y$, $i.e.$ $\Delta T_y (H) = \left[\Delta T_y (T,H) - \Delta T_y (T,-H)\right]/2$.

\subsection{ARPES Measurements} 
All ARPES data were collected at the MAESTRO beamline 7 of the Advanced Light Source using the $\mathrm{\mu}$ARPES endstation with spot size typically 20 \micron, energy resolution 10-20 meV, and sample temperature 10 K.  Samples were cleaved \textit{in situ} and measurements conducted at pressure $P < 3 \times 10^{-11}$ Torr.  Before ARPES each sample was surveyed spatially to identify $H$ and $T$-terminated surfaces according to their characteristic core level spectra.  ARPES spectra were collected using a Scienta R4000 analyzer with customized deflector elements so that for each ARPES map, the X-rays and sample are in a fixed geometry.

{Typically, cleaved samples expose both $T$- and $H$-terminated regions whose typical size ranges from submicron to some 10's of microns.  Therefore the employed spatial resolution of 20 \micron\  was appropriate to conduct comparisons between $H$ and $T$ terminated surfaces. We typically located these regions through rapid scans of the core levels which showed characteristic fingerprints as discussed in the supplemental section.}

\subsection{STM Measurements}

STM measurements were performed using a Unisoku STM, using chemically etched and annealed tungsten. Spectra were acquired using a standard lock-in technique at a frequency of 907 Hz. The samples are cleaved in the plane perpendicular to the $c-$axis at temperatures of about 90 K in UHV conditions ($P\sim$ 5E-10 torr) and then immediately inserted into the head of the scanning tunneling microscope.

\clearpage

\section{SUPPLEMENTARY MATERIALS}

\subsection{Temperature-dependent Hall resistivity}
\label{sec:hall}

\begin{figure}[t]
\centering
\includegraphics[width = 0.65 \linewidth]{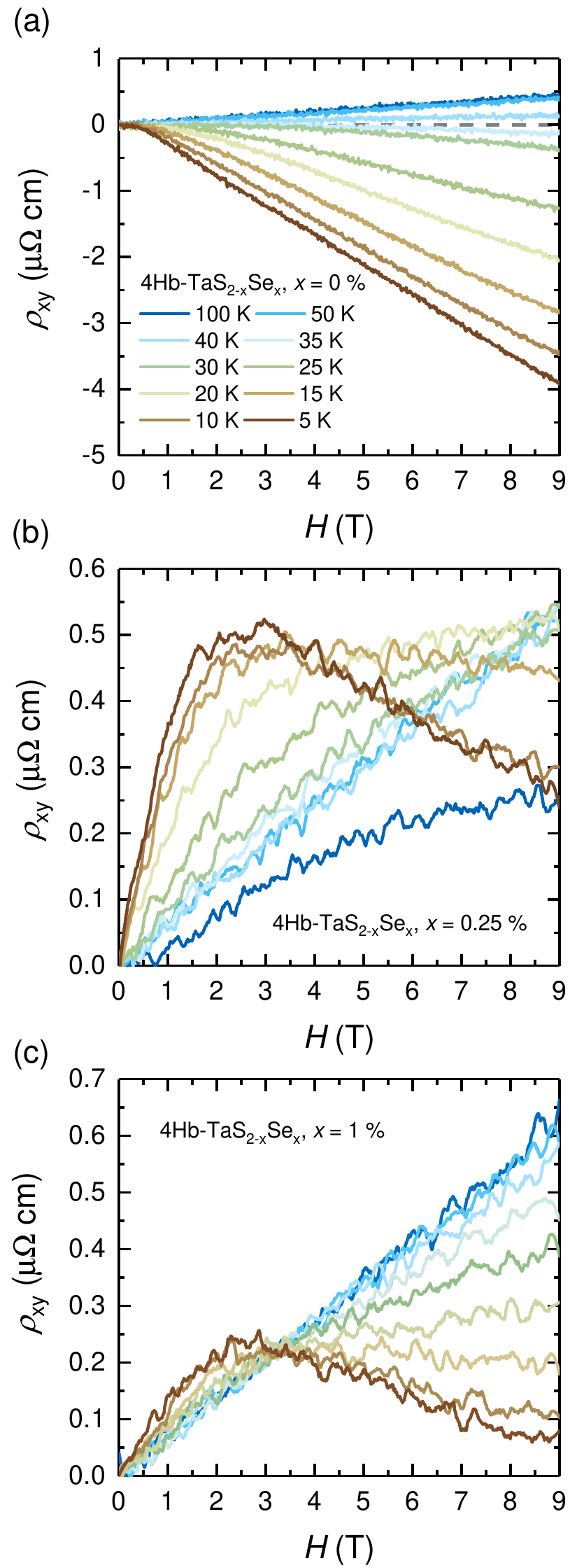}
\caption{
Field dependence of in-plane electrical Hall resistivity $\rho_{xy}(H)$ for \hb~(a) $x=0 \%$, (b) $x=0.25 \%$, and (c) $x=1 \%$ samples measured at several selected temperatures up to $H$ = 9 T. The $x=0 \%$ sample shows nearly a linear-in-field dependence in $\rho_{xy}(H)$, while the $x=0.25 \%$ and $x=1 \%$ samples show a non-linear field dependence at low temperature and gradually evolve toward a linear-in-field behavior as temperature increases. $\rho_{xy}(H)$ clearly shows a sign change from negative to positive around 35 K in the $x=0 \%$ sample, which is tied to the low-temperature CCDW transition, while it remains positive within the measured temperature range for the $x=0.25 \%$ and $x=1 \%$ samples.
}
\label{figS2}
\end{figure}

The field dependence of in-plane electrical Hall resistivity $\rho_{xy}(H)$ for \hb~$x=0 \%$, $x=0.25 \%$, and $x=1 \%$ samples measured at selected temperatures are plotted in Fig. \ref{figS2}. For $x=0 \%$ sample, $\rho_{xy}(H)$ is linear-in-field above $H$ $\sim$ 1 $T$ at all measured temperatures and has a non-linear field dependence below $\sim$ 1 T. For $x=0.25 \%$ and $x=1 \%$ samples, $\rho_{xy}(H)$ shows a stronger non-linear field dependence below $\sim$ 30 K, clearly suggesting multiple bands participate in $\rho_{xy}(H)$. $\rho_{xy}$ of 0 \% sample shows a sign change from negative to positive as temprerature increases above 35 K \cite{Gao2020}, consistent with a CCDW transition that happens around 30 K (as shown in Fig. \ref{Fig:transport_mag} (c)). Above 35 K, $\rho_{xy}$ is dominated by a hole band, while it is dominated by an electron band below 35 K. Compared to the 0 \% sample, 0.25  \% and 1 \% samples are dominated by the hole band across the whole measured temperature range. The change in the sign and field dependence of $\rho_{xy}$ reveal that the underlying CCDW phases - modulated by Se concentration - strongly alters the Fermi surface structure and related band parameters.

\subsection{Chemical origin of Se substitution for core-level spectroscopy}
\label{sec:corelevel}
\begin{figure}[t]
\centering
\includegraphics[width = 1.0 \linewidth]{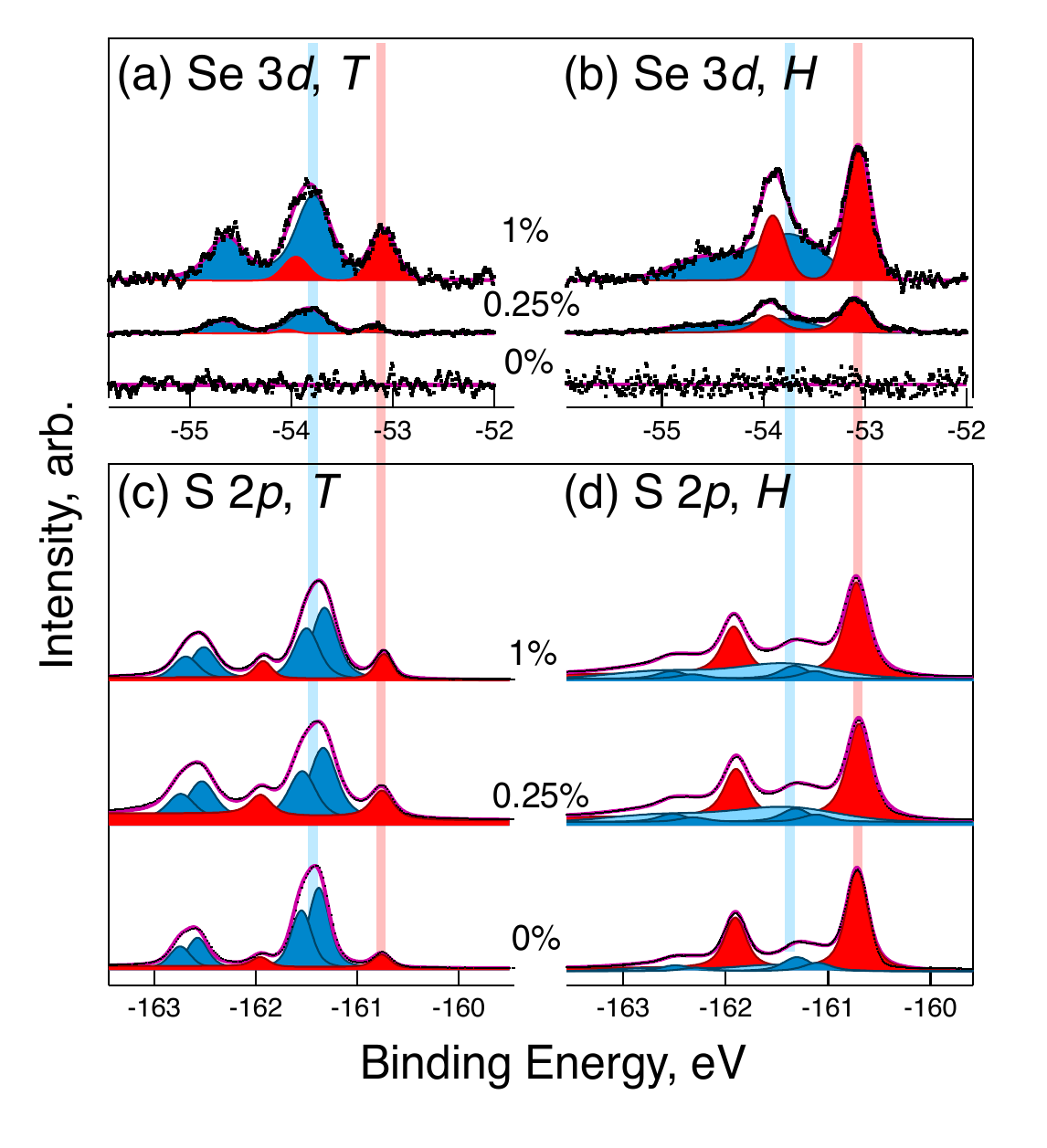}
\caption{
(a,b) Se 3$d$ and (c,d) S 2$p$ x-ray photoelectron spectra for $T$- and $H$-terminated surfaces respectively, and for Se concentrations of 0\%, 0.25\%, and 1.0\%.  The spectra are comprised of spin-orbit split doublets (red or blue) each of which represents a distinct atomic site near the surface.  The 0\% sample was of the insulating type designated \zpi.  The Se spectra are presented after subtraction of a smooth background for clarity.  The binding energy ranges (a-d) are plotted on the same scales, so that the leading binding energies for H and T core level peaks may be directly compared for Se 3$d$ and S 2$p$ electrons.
}
\label{XPS}
\end{figure}

Fig.\ \ref{XPS} shows x-ray photoelectron spectroscopy (XPS) results for Se and S atoms in \hb. Although the concentration is low, it is possible to observe Se $3d$ core electrons for both $T$- and $H$-terminated surfaces in Fig.\ \ref{XPS}(a,b), respectively.  These spectra consist of two spin-orbit split doublets, whose intensities are proportional to their proximity to the surface, owing to the short escape depth of the electrons.  Thus in (a) we assign the stronger, blue doublet to Se in the outermost surface layer with $1T$ termination, while in (b) the stronger, blue doublet is assigned to $1H$ termination.

The energy shift between the $H$ and $T$ is due to their chemical state difference and can be explained by the charge transfer from $T$ to $H$ layers.  However, there are also final state relaxation contributions to the absolute binding energies that relate to the local electronic screening in each layer.  The data also clearly show a proportional growth of spectral weight as Se concentration increases, verifying our stated Se concentrations. 

The simple 2-component fitting model employed hides the complexity of the distribution of chemical shifts expected from the CDWs in each layer through the inclusion of Gaussian broadening.  The buried $T$-layer's Se doublet in (b) is much broader than in the exposed $T$ layer seen in (a). This indicates a larger spread of charge density among atomic sites in the buried layers, either due to enhanced CDW strength, or due to enhanced interactions with $H$ layers above and below.  Further investigation is needed to understand this.

Fig.\ref{XPS}(c,d) show corresponding spectra for S $2p$.  Like Se $3d$, these also show a partitioning of the spectrum into (to lowest order) two doublets corresponding to $T$- and $H$-layer S atoms with their intensities ordered by their surface proximity.  From this we can conclude that the Se and S atoms sit in chemically identical states, with identical chemical shifts.  This confirms that the Se occupies S sites, and dopes the $T$ and $H$ layers roughly equally.

Compared to Se, the S statistics are much better and we could discern additional fine structure, such as a small splitting of the $T$-layer's peaks. Furthermore, the spectra are well-fitted when the $H$ layers include a tail towards high binding energy, modelled using the Shirley function.  This reflects the fact that $H$ layers are metallic, with the tail representing energy loss of the outgoing electrons to Fermi liquid excitations. For the $H$ surfaces, we found, as for Se, that the buried $T$ S atoms have a broader energy distribution.  This was represented by a broad doublet (light blue) combined with two sharper peaks (darker blue).

\subsection{H-layer Fermi surfaces and charge-transfer analysis}
\label{sec:Hlayer}

\begin{figure}[htb!]
\centering
\includegraphics[width=\columnwidth]{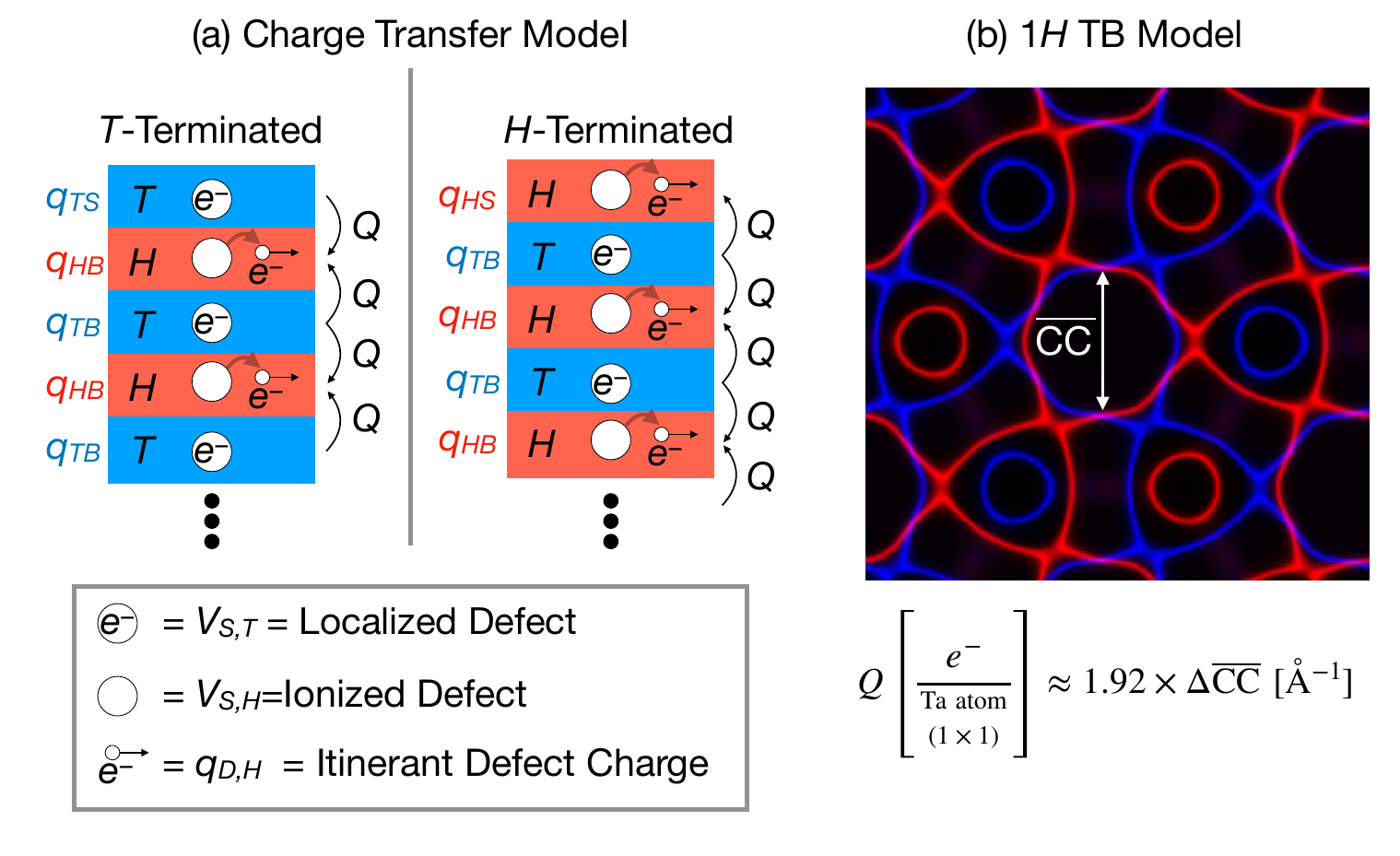}
\caption{Charge transfer model and \cc\ metric. (a) Schematic of the layer-resolved charge transfer in \Fhb\ for $T$- and $H$-terminated surfaces. Each \Tl\ donates charge $Q$ per interface to each adjacent \Hl; surface layers receive $Q$ while bulk layers receive $2Q$ due to two adjacent interfaces. The layer fillings $q_{TS}$, $q_{TB}$, $q_{HS}$, and $q_{HB}$ are defined relative to the isolated-layer reference and are summarized in Table~\ref{tab:tableCC}. When present, negative or ionized S vacancies are indicated by large white circles.  (b) Calculated spin-up (red) and spin-down (blue) Fermi contours of the 1$H$ TB model, with the \cc\ metric indicated. An empirical rule for estimating the charge transfer $Q$ is included.}
\label{Fig:qxfer}
\end{figure}

Fig.\ \ref{Fig:ARPES}(c,d) show the Fermi surfaces of the \Hs\ layers for the different sample classes. These Fermi surfaces are well-reproduced by the TB calculation for monolayer 1$H$. They are qualitatively similar to those of the buried \Hl s in (a-c), but are stronger and sharper due to the lack of attenuation through overlaying layers, and differ in details such as the topology around a saddle point Van Hove Singularity (VHS, marked V in the figure) and overall dimensions. These differences are understood within a simple charge transfer model illustrated in Fig.\ \ref{Fig:qxfer}, in which each \Tl\ donates a fixed negative charge $-Q$ to its adjacent $H$ layers. Due to the surface termination, the surface $T$ and $H$ layers will have net charge $\pm Q$ while deeper layers will have net charge $\pm 2Q$. Therefore the charge transfer difference between surface and subsurface $H$ determines $Q$ and the overall charge state of all layers.

Density Functional Theory (DFT) calculations for the SOD reconstruction predict the \Tl\ to host a narrow half-filled flat band derived from \dzz\ orbitals at \EF. The flat band holds exactly 1\,\el\ per SOD cluster (13 Ta atoms), or equivalently $1/13 \approx 0.077$\,\el\ per Ta atom in the 1$\times$1 unit cell. Quantifying the charge transfer between the $T$ and $H$ layers is therefore important for determining the occupancy of this band. To estimate the charge content in our bulk layers, we begin by estimating the interlayer charge transfer $Q$ using Luttinger's theorem, by comparing the Luttinger area of surface and buried $H$ layer Fermi contours.  Previously, $Q$ was estimated by comparing the \Hl\ Fermi surface of \Fhb\ to that of \Thb\cite{Almoalem2024ARPES}. In contrast, our measurement compares two side-by-side spots on the same \Fhb\ crystal — one H-terminated (surface \Hl) and one T-terminated (subsurface \Hl) — which eliminates random measurement errors, and systematic errors from sample quality, sample-to-sample differences in band structure, lattice constant, and work function.

To quantitatively analyze the $H$-layer Fermi surfaces, we fitted the experimental data using a tight-binding (TB) model based on the three-band parametrization of Liu \textit{et al.}~\cite{Liu2013}. The fitting was performed using the $H$-terminated surface of the 0\% Se sample measured at $T=46$ K, which is nominally above the low-temperature $3\times3$ CDW transition of the $H$ layer in \Fhb. The model does not include the $3\times3$ or \rthirteen\ reconstructions and is therefore used only to describe the underlying unreconstructed $H$-layer Fermi surface and estimate charge transfer on a per-Ta basis relative to the ($1\times1$) unit cell.

In Fig.~\ref{Fig:ARPES}(a,d), the TB models have been adjusted to the different charge levels in the (a) subsurface and (d) surface $H$ layers. This amounts to a relative shift of \EF\ in (d) being 16 meV lower than in (a). This energy shift is corroborated by a similar bandwidth change \dW\ in the bandstructure slices, which is similarly narrower in (d) than in (a). Furthermore, the shape of the bands near point V shows that in (a) the VHS is nearly at \EF\ while in (d) the VHS has been pushed above \EF. However, calculation of the charge transfer based on the measurement of \dW\ or the position of V is complicated because the bottom of the band and the VHS are influenced by the CDWs that are not included in the TB model. Therefore, we optimize the fits using features at the Fermi surface that are least affected by the CDWs. 

Namely, due to mirror symmetry in the central plane of the \Hl, the out-of-plane spins are completely decoupled eigenstates, resulting in two bands of opposite spin having six symmetry-protected crossings C. The bands at C are typically strong and sharp, are protected from splitting, and are not strongly affected by the CDWs. Denoting the distance between opposite C points as \cc, we derive from the TB model the rate of charge accumulated by the \Hl\ for a given change in \cc. Between Fig.~\ref{Fig:ARPES}(a) and Fig.~\ref{Fig:ARPES}(d), we determine $\Delta$\cc\ $= 0.011$~\AA$^{-1}$, corresponding to $Q = 0.021$\,\el\ per Ta atom for the \zpi\ sample. Using the algebra summarized in Table~\ref{tab:tableCC}, the bulk \Tl\ filling is $1 - 2\times 13Q = 0.46$\,\el/SOD for \zpi\ samples, rising to $0.70$\,\el/SOD for \opm\ samples. In both cases the flat band retains the majority of its charge, confirming that the \Tl\ is not depopulated.  Note that for the \zpi\ sample, we are assuming that there is no defect-induced charge transfer as discussed for the \zpm\ sample in Section \ref{sec:0M}. This is established by the lack of observable S vacancies in Fig.\ \ref{Fig:STM} (a,d). 

By focusing on a convenient metric--the distance between protected band crossings C in Fig.\ \ref{Fig:qxfer}(b), and by analysis of our TB model's density of states \vs\ energy and band slope at the crossing points, we determine the empirical relationship:
\begin{equation}
Q\,\left[\mathrm{e}^-/\mathrm{1\times1}\right] = 1.92 \times \Delta\ccmath\,[\mathrm{\AA}^{-1}],
\label{eq:rulethumb}
\end{equation}
where $\Delta$\cc\ is the change in distance between opposite band crossings C. The coefficient 1.92 is derived from the TB band slope $dE/dk = 1.08$\,eV\,\AA\ at the crossing and the TB density of states $dN/dE = 3.55$\,\el/eV per Ta atom at \EF, giving $\Delta k \times (dE/dk) \times (dN/dE) = (\Delta\ccmath/2) \times 1.08 \times 3.55 = 1.92\,\Delta\ccmath$. To assess what fraction of the flat band in the \rthirt-reconstructed surface \Tl\ is transferred, we multiply $Q$ by 13 to convert to e$^-$/SOD; complete depopulation of the flat band would correspond to $13Q = 1$\,\el/SOD. These values of \cc\ and $Q$ are summarized for \zpi, \opm, and \zpm\ samples in Table~\ref{tab:tableCC}. These confirm the presence of charge transfer from $T$ to $H$ layers. However, the values of $Q$ are substantially smaller than those reported earlier \cite{Date2025,Almoalem2024ARPES}, with values in the range $0.15 < 13Q <0.4 \ll 1$, demonstrating that the flat band is far from completely depopulated by charge transfer alone, consistent with the residual \dzz\ spectral weight observed in the \zpi\ samples (Sec.~\ref{sec:polarization}).  

The smallest charge transfer and highest $T$ flat band occupation were observed for the \opm\ sample.  This result suggests a small charge-doping effect of Se vacancies, which would act to reduce the interlayer charge transfer $Q$.  A similar trend with Se doping was previously observed for much higher Se concentrations\cite{Geng2024}.

\begin{table*}[htb!]
\centering
\caption{Extracted charge transfer parameters and layer fillings for the three sample classes. $Q$ is the charge transferred per 1$\times$1 unit cell across a single 1$T$/1$H$ interface; $q_{d,H}$ is the additional \Hl\ charge from native S vacancies. The 1$T$ filling rows are per SOD cluster (13$\times$13 unit cell), where the half-filled flat band holds 1\,e$^-$/SOD. See Sec.~\ref{sec:Hlayer} and Sec.~\ref{sec:0M} for full details.}
\label{tab:tableCC}
\begin{tabular}{l l c c c c}
\hline\hline
\textbf{Parameter} & \textbf{Symbol (Formula)} & \textbf{Units}
    & \textbf{0\%I (Pristine)} & \textbf{1\%M (Se-doped)} & \textbf{0\%M (Defective)} \\
\hline
\cc\ distance (1$T$-surface) &                               & \AA$^{-1}$         & 1.120  & 1.115  & 1.031  \\
\cc\ distance (1$H$-surface) &                               & \AA$^{-1}$         & 1.131  & 1.121  & 1.046  \\[4pt]
Interface transfer            & $Q$                           & e$^-$/1$\times$1   & 0.021  & 0.012  & 0.030  \\
Bulk 1$H$ charge transfer     & $q_{HB}$ ($=2Q$)             & e$^-$/1$\times$1   & 0.042  & 0.023  & 0.060  \\
H-layer defect charge         & $q_{d,H}$                    & e$^-$/1$\times$1   & 0      & 0      & 0.20   \\[4pt]
Surface 1$T$ filling          & $q_{TS}$ ($=1-13Q$)          & e$^-$/13$\times$13 & 0.73   & 0.85   & 0.61   \\
Bulk 1$T$ filling             & $q_{TB}$ ($=1-2\times13Q$)   & e$^-$/13$\times$13 & 0.46   & 0.70   & 0.22   \\
\hline\hline
\end{tabular}
\end{table*}


\subsection{Metallic 0\% Se samples and the role of native disorder}
\label{sec:0M}

\begin{figure}[b]
\centering
\includegraphics[width = 1.0\columnwidth]{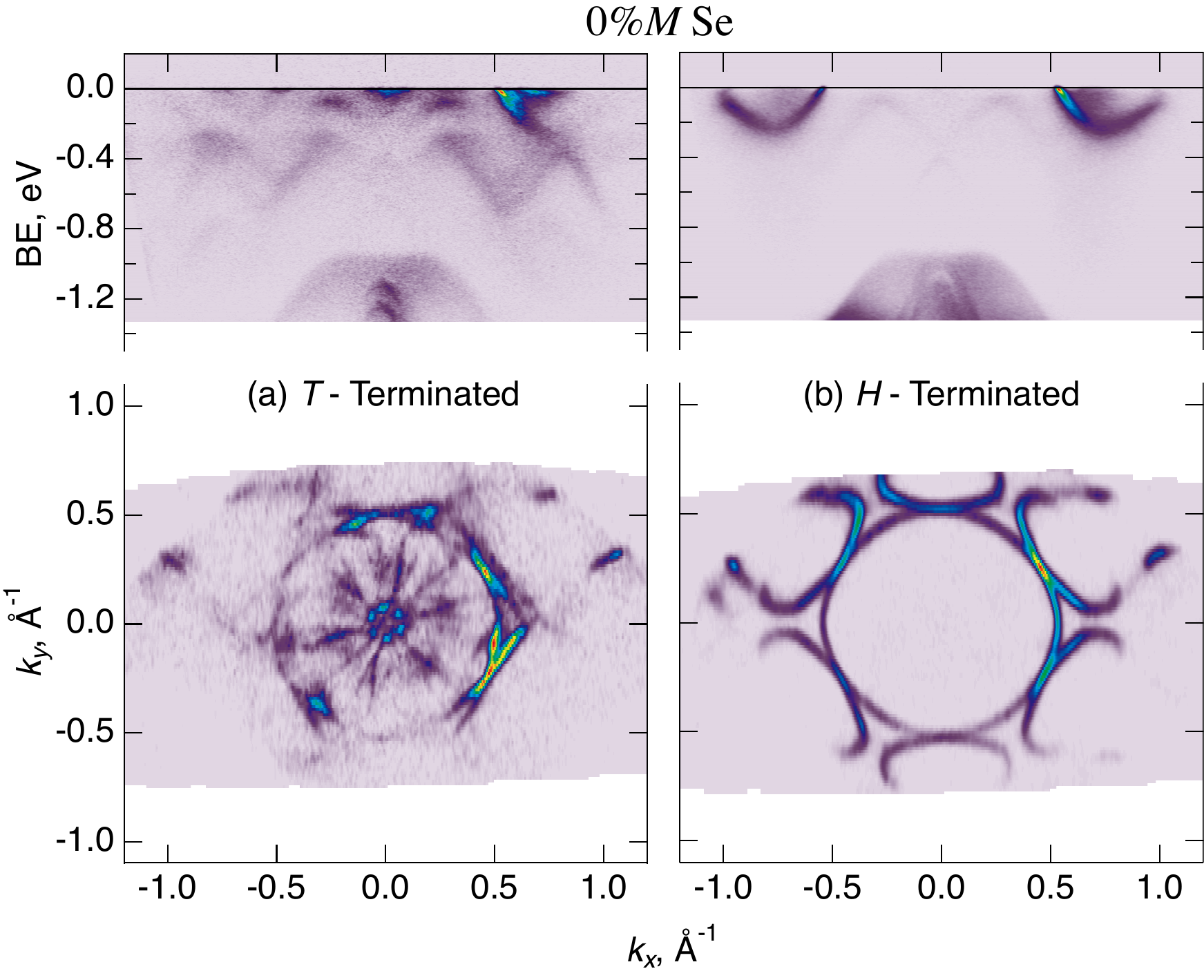}
\caption{{\bf ARPES evidence for a delocalization transition in \hb\ with no Se.} (a,b) Data is the same as that collected Figure \ref{Fig:ARPES} of the main text for $T$- and $H$-terminated surfaces, but on a sample with 0$\%$ Se substitution containing intrinsic disorder, presumably due to S vacancies. While such samples were less commonly found, it is likely that quench cooling or cleaving conditions can create enough disorder on the surface as to delocalize the flat band on the \Tl.  
}
\label{Fig:ARPESZM}
\end{figure}

Fig.\ \ref{Fig:ARPESZM}(c) shows a third, anomalous class of samples that have no Se, and yet display the metallic windmill states in the \Ts. Such \zpm\ samples are a minority of the 0\% samples examined. We have found by cleaving samples multiple times that the metallic or insulating character of the $T$ layer is consistent throughout a given crystal, therefore the metal/insulator distinction is probably due to small variations of growth conditions affecting each entire crystal and not due to inhomogeneities within a given crystal.

The possibility of seeing windmill states without Se doping suggests that this transition can occur also for other sources of disorder. It also reconciles the variety of different spectra in the literature claimed for 0\% Se doping which have seen either surfaces similar to our \zpi\ samples \cite{Almoalem2024ARPES} or surfaces similar to our \zpm\ samples \cite{Date2025}. What sets our results apart is that we have baseline insulating \zpi\ samples for quantitative comparison. The Fermi surface dimensions in both the surface and next-deep \Hl s show a significant accumulation of electrons in the Ta $d$-bands of the \zpm\ samples, evidenced by both qualitative assessment of topology (wider dog's bones) and quantitative analysis (reduced \cc\ distance) compared to the standard \zpi\ samples. Since XPS uncovered no dopants or contaminants (see Sec.~\ref{sec:corelevel}), the natural explanation for this charge accumulation is S vacancies. Based on the cross-sample \cc\ comparison and Eq.~\ref{eq:rulethumb}, we estimate that \Hl s in \zpm\ samples carry an excess of around 0.20 electrons per Ta atom (1$\times$1 unit cell) relative to the \zpi\ samples, corresponding to an S vacancy density of around 5\% (each S vacancy donates 2 electrons, with 2 S sites per Ta).

S vacancies can in principle add additional defect charge terms $q_{d,1T}$ or $q_{d,1T}$ to the $T$- or $H$-layers. However, it has been argued that this excess charge resides in the \Hl\ rather than the \Tl. First-principles calculations have shown that S vacancies in the \Tl\ (V$_\mathrm{S,1T}$) distort the local CDW structure and suppress the flat band locally without doping the surrounding \Tl\ \cite{yang_nanoscale_2025}, so V$_\mathrm{S,1T}$ defects contribute negligibly to the bulk electron count; accordingly we set $q_{d,T} = 0$ in our model. The present data provide the first experimental evidence that S vacancies in the \Hl\ (V$_\mathrm{S,1H}$) directly accumulate charge in the \Hl: no other plausible source of excess electrons such as impurity elements is apparent in our XPS survey spectra, and the reduced size \Hl\ central Fermi surface of the \zpm\ samples cannot be explained by the small intrinsic charge transfer $Q$ alone. This is consistent with the theoretical expectation that V$_\mathrm{S,1H}$ defects locally suppress the $T\rightarrow H$ charge transfer, returning electrons to the system and raising the net \Hl\ filling \cite{yang_nanoscale_2025}. The intrinsic $T\rightarrow H$ charge transfer $Q$ is found to be $\sim 40$\% larger for \zpm\ compared to the \zpi\ samples (Table~\ref{tab:tableCC}).  This shows that while local charge transfer  is suppressed near defects, \cite{yang_nanoscale_2025}, it is not globally suppressed by defects; the bulk 1$T$ filling is similar for \opm\ and \zpm samples.
d
The quantitative estimates of the layer-dependent filling in Table\ \ref{tab:tableCC} clearly establish that the amount of charge in the \Tl\ flat band is not strongly correlated with whether the \Tl\ is in the localized or metallic state. When S-site defects are introduced, the collapse of the flat band into a metallic phase is inconsistent with charge filling in the \Tl\ band—it becomes heavily populated with Se substitution in the \opm\ samples, yet heavily depleted by native sulfur vacancies in the \zpm\ samples. Because the delocalized, metallic state emerges regardless of whether the absolute band filling is higher or lower relative to the pristine baseline, charge transfer cannot be the sole driver of the transition. Instead, the delocalization is driven by the structural disorder introduced by S-site defects. This defect-driven interpretation is supported by recent STM observations \cite{Geng2024} showing that at higher Se concentrations, spatially inhomogeneous flat-band filling emerges in the \Tl, with distinct electron-filled and electron-void SOD clusters whose ratio tracks the effective doping level. 

\subsection{Polarization-dependent ARPES of occupied \dzz~states in insulating 1$T$ layers}
\label{sec:polarization}
\begin{figure}
\centering
\includegraphics[width = 1 \linewidth]{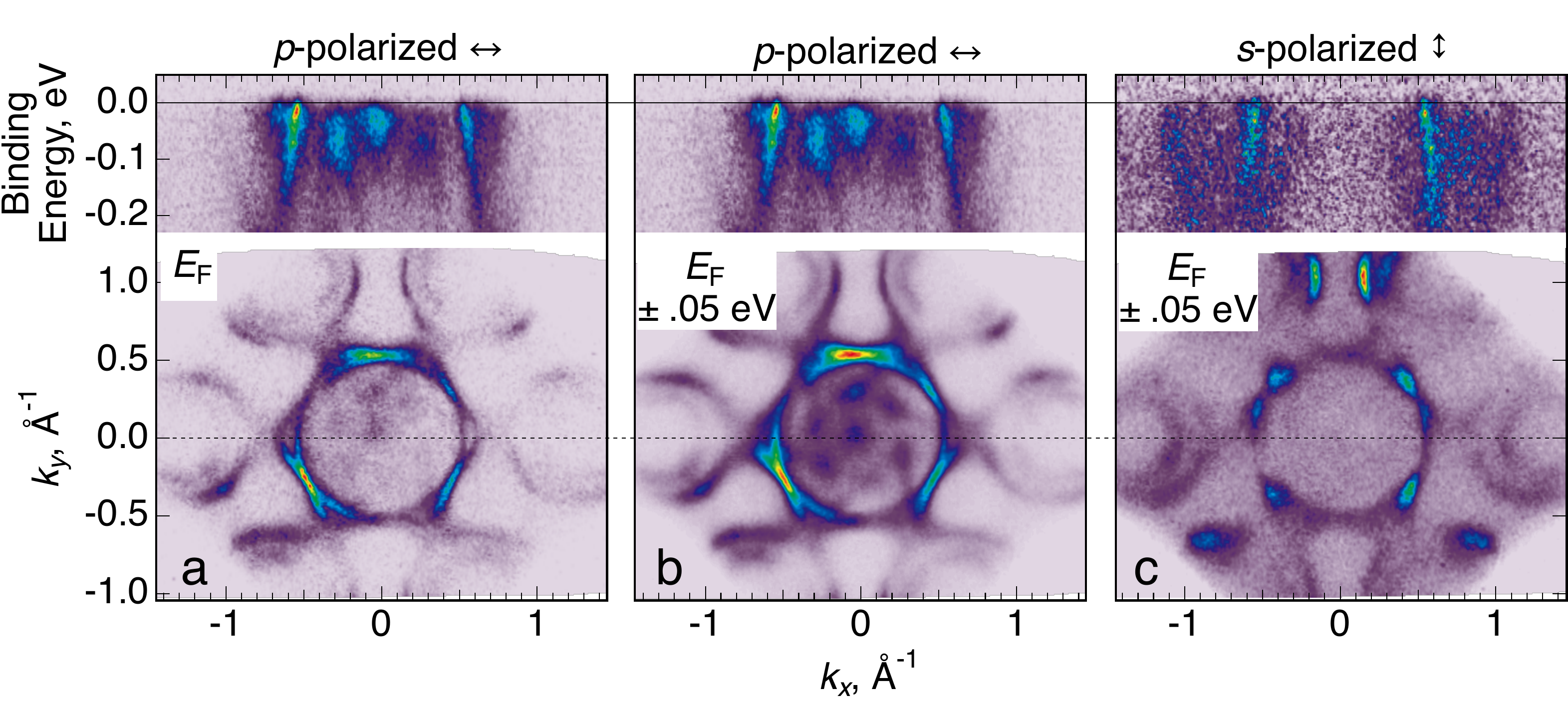}
\caption{{\bf ARPES evidence for occupied \dzz orbitals in insulating $T$ surface layers of 0\% Se}.  (a) Bandstructure along \ky=0 (upper) and Fermi surface (lower) of \zpi\ sample, collected at 102 eV with $p$-polarized light.  (b) same, with integration window $\pm 50$ meV to highlight the concentration of diffuse states just below \EF.  (c) same as (b) but collected with $s$-polarized light.  The states near \G\ are suppressed due to \dzz\ orbital character, while the remaining states from the underlying \Hl\ are preserved due to their partial \dxy character. 
}
\label{Fig:ARPESPOL}
\end{figure}

The bound states associated with the insulating 1$T$-layer state in the \zpi\ samples are visible as broad spectral features and a high diffuse background within $\sim 100$ meV below \EF, seen in Fig.\ \ref{Fig:ARPES}(a) for 72 eV, as well as in Fig.\ \ref{Fig:ARPESPOL}(a) for 102 eV light.  These states resemble those seen in \cite{Almoalem2024ARPES}, but here we report that these states have little or no spectral weight at \EF.  The momentum distribution of these states is made clear with a larger energy integration window, see Fig.\ \ref{Fig:ARPESPOL}(b) showing the  rotation of the features consistent with the chiral SOD reconstruction.

In order to investigate the orbital character of these \Tl\ states, we switched the x-ray polarization from $p$-polarized (light polarized in the scattering plane) to $s$-polarized (light polarized perpendicular to the scattering plane, and parallel to the \ky\ direction).   According to the basic theory of ARPES \cite{moser_2017}, orbitals that are even with respect to the $xz$ scattering plane will be visible in $p$-polarization but suppressed in $s$-polarization, namely \dzz, \dxz, and \dxx. The opposite polarization dependence holds for \dxy\ and \dyz\ orbitals that are odd with respect to the scattering plane. The low-energy states of the \Tl\ are well-known to be \dzz-derived, centered on \G\ and localized to the central Ta atom of the SOD clusters, while the low-energy states of the \Hl\ are located away from \G\ and are derived from a hybrid combination of even and odd \dxy, \dxx, and \dzz\ orbitals, the latter contributing weakly away from \G.  Thus we expect the \Tl\ states, if any, to be present in $p$ but completely absent in $s$ polarization, while the \Hl\ states to be partially present in both $p$ and $s$, depending on the details of the hybridization. 

Fig.\ \ref{Fig:ARPESPOL} compares the ARPES Fermi Surface and low-lying bandstructure for (a,b) $p$-polarization and (c) $s$-polarization.  These states show that all traces of the diffuse states near \G\ are extinguished in $s$-polarization, consistent with the expectation that these are \dzz-derived states associated with the SOD Ta atoms.  The remaining states in (b) are retained in (d), but their intensities are suppressed according to the relative contribution of even \dzz/\dxx orbitals (that are suppressed in (c)) and \dxy~(that are not suppressed in (c)).  These results strongly support the assignment of the diffuse states to \dzz\ orbitals of the central Ta in the SOD clusters.

\subsection{STM evidence for superconductivity in fragmented CDW regions}
\label{sec:STM}

Additional STM measurements are shown in Supplementary Fig.~\ref{Fig:STM_supp}. Each row in Supplementary Fig.~\ref{Fig:STM_supp} corresponds to a different local region of the pristine \Fhb\ sample. 

\begin{figure*}[t]
\centering
\includegraphics[width = 0.85\textwidth]{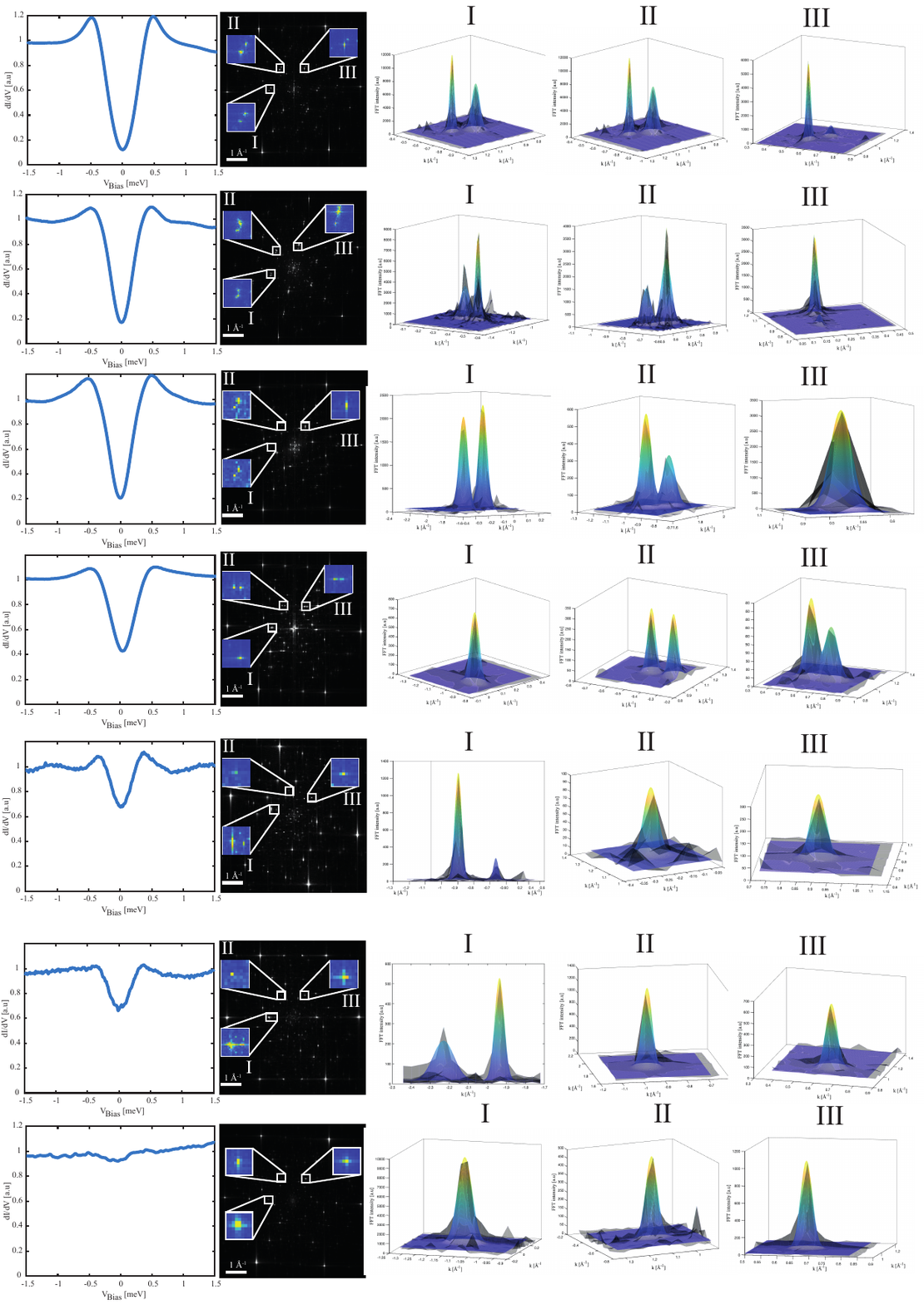}
\caption{
Additional STM measurements showing the correlation between CDW fragmentation and superconducting gap formation at different locations of the 1H layer in pristine \Fhb samples.
Each row shows (left) the local tunneling spectrum $dI/dV$, (middle) the corresponding Fourier transform of the CDW order, and (right) enlarged views of the CDW Bragg peaks labeled I, II, and III.
Regions with stronger CDW peak splitting, which are indicative of more fragmented SOD order and enhanced domain-wall density, exhibit larger superconducting gaps.
In contrast, regions with sharper and unsplit CDW peaks show strongly reduced or absent superconducting gaps.
}
\label{Fig:STM_supp}
\end{figure*}

\subsection{Thermal Conductivity}
\label{sec:kappaxx}

\begin{figure}[t]
\centering
\includegraphics[width = 0.9 \linewidth]{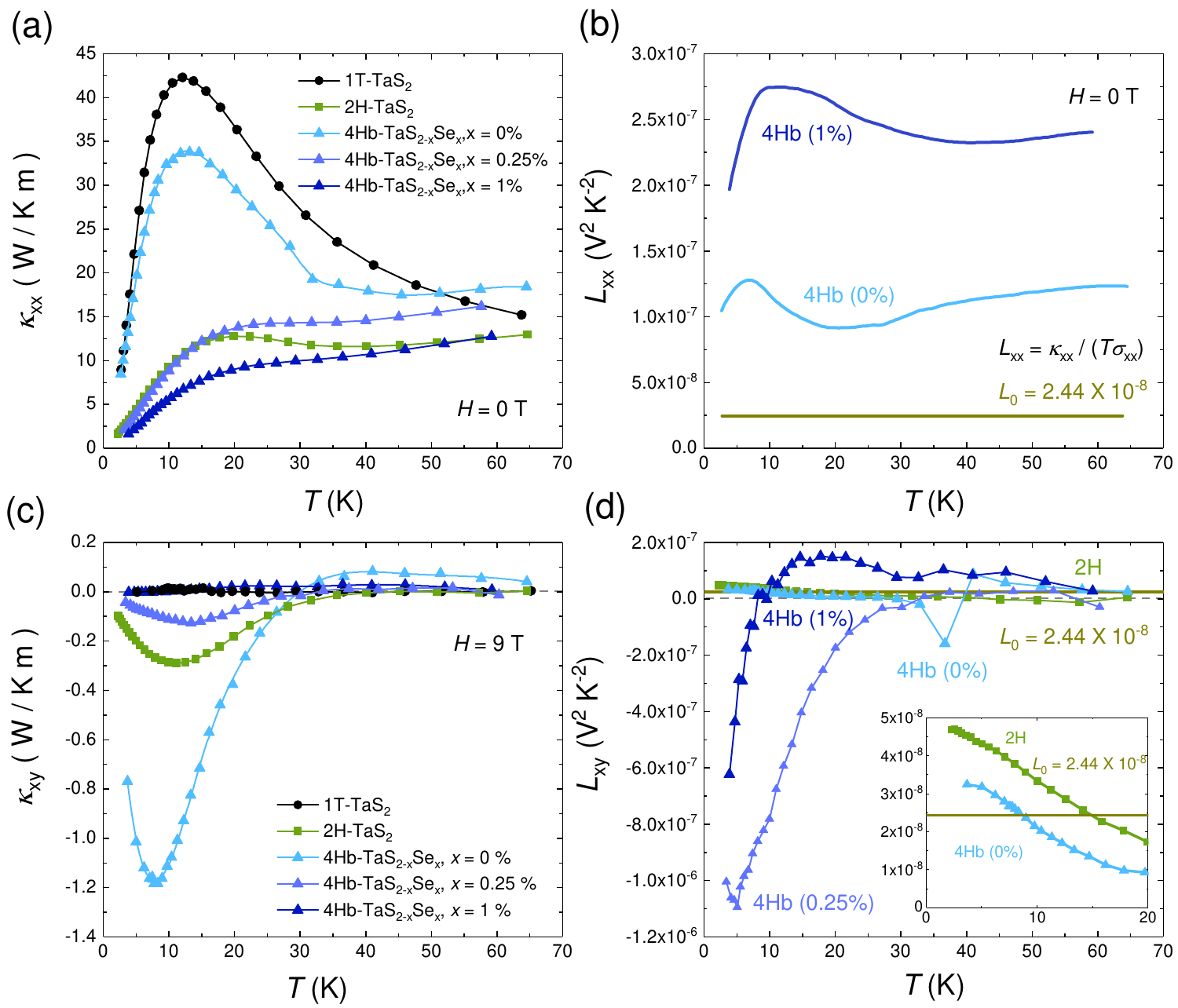}
\caption{{\bf Thermal conductivity, thermal Hall conductivity, and quenched disorder in \hb.}
(a) Comparison of thermal conductivity \Kxx~of 1T-, 2H- and \Fhb~structures, illustrating a dramatic drop in phononic thermal conductivity peak and thus an increase in phonon scattering when 1\% Se is added in \Fhb. (b) Calculated Lorenz number $L_{xx} = \kappa_{xx}/(T\sigma_{xx})$ in \hb~$x$ = 0\% and 1\% samples. Lorenz number expected of a conventional metal $L_{0}$ is shown as the horizontal dark yellow line.  The deviation from the Wiedemann-Franz law becomes more pronounced in the $x$ = 1\% sample, suggesting the electrons are much more strongly scattered due to enhanced interactions. (c) Thermal Hall conductivity \Kxy~of 1T-TaS$_{2}$, 2H-TaS$_{2}$, \hb~$x$ = 0\%, $x$ = 0.25\%, and $x$ = 1 \% samples as a function of temperature in a magnetic field of $H = 9$~T.
In \hb~, \Kxy~is mainly contributed by electrons. The amplitude of \Kxy~is strongly suppressed as Se concentration increases, which further illustrates the enormous effect of electron scattering on adding Se. \Kxy~is almost completely suppressed at $x$ = 1\%, similar to the 1T-TaS$_{2}$ case where the electrons are completely localized. (d) Calculated Lorenz number $L_{xy} = \kappa_{xy}/(T\sigma_{xy})$ in 2H-TaS$_{2}$, \hb~$x$ = 0\%, $x$ = 0.25\%, and 1\% samples in a magnetic field of $H = 9$~T. The WFL is mostly fulfilled in 2H-TaS$_{2}$ and pristine \hb (as shown in the main panel (d) and inset zoom-in at low temperature), while it is strongly violated in the Se substituted \hb~samples.}
\label{Fig:4Hbtherm}
\end{figure}

 To understand how lattice disorder connects to the apparent electronic emergence, we also study the thermal conductivity and thermal Hall conductivity in the $1T, 2H$ and \hb~systems. Such measurements can be used to analyze the Wiedemann-Franz law, which states that at sufficiently low temperature all metals should satisfy $\kappa/\sigma T = L_0$, where $L_0 = 2.44\times 10^{-8}$V$^2$/K$^2$ is the Lorenz number. As stated in the main manuscript, the law states that for normal metals the majority carriers of charge and entropy should be electrons at temperatures where phonons are frozen out. Deviations from the Wiedemann-Franz law can therefore indicate  the presence of neutral excitations, such as phonons that do not carry charge, yielding much larger values of $L_0$, as well as unconventional electronic transport arising from strong scattering or incoherent quasiparticle dynamics. 

Figure \ref{Fig:4Hbtherm} (a) shows the evolution of the thermal conductivity for $1T-, 2H-$ and \hb~series, where phonons are the dominant heat carriers \cite{Nunez1985}. The features are consistent with the temperature dependence of a conventional phonon-dominated \Kxx, in which increasing thermal excitation of phonons initially enhances the thermal conductivity until higher-momentum states are populated and Umklapp scattering sets in, suppressing \Kxx~and producing the characteristic phonon peak \cite{Chen2022}. In \Fhb, the charge density wave transition at $\sim 30$ K will also create additional scattering of phonons from electrons. 

The thermal conductivity of the pristine \Fhb~is second only to the $1T-$ structure, indicating the pristine \Fhb~has a slightly higher lattice disorder than $1T$-TaS$_{2}$. As the Se concentration grows, the thermal conductivity is suppressed in harmony with the increasing electrical resistivity due to an increased scattering of both phonons and electrons by impurities. The calculated Lorenz ratio $L_{xx} = \kappa_{xx}/(T\sigma_{xx})$, shown in Fig. 4(b), exceeds the Sommerfeld value $L_{0}$. Deviations of $L_{xx}$ from $L_{0}$ reflect the relative efficiency of heat transport compared to charge transport. Although both phononic and electronic contributions to \Kxx~are reduced upon Se substitution, the deviation of $L_{xx}$ from $L_{0}$ becomes more pronounced in the $x$ = 1\% sample, indicating that disorder suppresses electrical conductivity more strongly than total thermal conductivity. These results are consistent with phonon-dominated heat transport, where disorder disproportionately degrades charge transport relative to heat transport.

As described in the main text, the thermal Hall effect suggests strong inelastic scattering of the electrons as a function of disorder. As shown in Figure \ref{Fig:4Hbtherm}(c), the pristine \Fhb\ system has a substantial thermal Hall signal, which can be attributed to the existence of high-mobility electrons. (Note that the thermal Hall signal in $1T-$ is almost absent, consistent with its insulating behavior.) Upon adding Se, the thermal Hall effect is strongly suppressed and vanishes nearly immediately at $x$ = 1\%, indicating a severe reduction of the contribution of mobile electrons in the heat transport channel. 

We believe this to be more than the effect of just disorder. The comparison to the 2H- system is quite instructive here. The 2H-TaS$_2$ has a thermal conductivity and thermal Hall conductivity about 3 to 4 times smaller than the pristine \Fhb, suggesting that the main difference is due to greater scattering in the 2H- structures. However, when Se is added, the \hb\, has a thermal Hall effect that is much more suppressed than the thermal conductivity, suggesting electrons are scattered much more strongly than would be expected from simple disorder effects. This implication is even clearer when studying the Lorenz ratio $L_{xy}$ from the thermal Hall effect (as shown in Fig. \ref{Fig:4Hbtherm}(d)). In the $x$ = 0\% system, $L_{xy}$  approaches $L_0$ at low temperatures, just as it does in 2H-TaS$_2$. However, upon adding Se, the Lorenz ratio strongly diverges from $L_0$, reaching nearly $L_{xy}/L_{0}\approx 45$ at $x$ = 0.25\% around $T_\mathrm{c}$. The deviation from the Wiedemann-Franz law suggests that the new electrons entering the Fermi surface form a highly incoherent and strongly scattered metallic state. Such scattering may arise from low-energy inelastic processes, including superconducting fluctuations near $T_\mathrm{c}$, as well as multiband effects.
\bibliography{reference}

\end{document}